\documentclass[%
 aip,apl,showpacs,superscriptaddress,numerical,amsmath,amssymb,floatfix,
reprint,natbib%
]{revtex4-1}

\usepackage{bm}%
\usepackage{graphicx}

\newcommand{\CF}{Co$_{\text{25}}$Fe$_{\text{75}}$}

\begin{document}

\title{High Spin-Wave Propagation Length Consistent with Low Damping in a Metallic Ferromagnet}%

\author{Luis Flacke}
\email{luis.flacke@wmi.badw.de}
\affiliation{\mbox{Walther-Meißner Institute, Bayerische Akademie der Wissenschaften, 85748 Garching, Germany}}
\affiliation{Physics Department, Technical University of Munich, 85748 Garching, Germany}

\author{Lukas Liensberger}
\affiliation{\mbox{Walther-Meißner Institute, Bayerische Akademie der Wissenschaften, 85748 Garching, Germany}}
\affiliation{Physics Department, Technical University of Munich, 85748 Garching, Germany}

\author{Matthias Althammer}
\affiliation{\mbox{Walther-Meißner Institute, Bayerische Akademie der Wissenschaften, 85748 Garching, Germany}}
\affiliation{Physics Department, Technical University of Munich, 85748 Garching, Germany}

\author{Hans Huebl}
\affiliation{\mbox{Walther-Meißner Institute, Bayerische Akademie der Wissenschaften, 85748 Garching, Germany}}
\affiliation{Physics Department, Technical University of Munich, 85748 Garching, Germany}
\affiliation{Nanosystems Initiative Munich, 80799 Munich, Germany}
\affiliation{\mbox{Munich Center for Quantum Science and Technology (MCQST), 80799 Munich, Germany}}

\author{Stephan Geprägs}
\affiliation{\mbox{Walther-Meißner Institute, Bayerische Akademie der Wissenschaften, 85748 Garching, Germany}}

\author{Katrin Schultheiss}
\affiliation{Helmholtz-Zentrum Dresden-Rossendorf, 01328 Dresden, Germany}

\author{Aleksandr Buzdakov}
\affiliation{Helmholtz-Zentrum Dresden-Rossendorf, 01328 Dresden, Germany}

\author{Tobias Hula}
\affiliation{Helmholtz-Zentrum Dresden-Rossendorf, 01328 Dresden, Germany}

\author{Helmut Schultheiss}
\affiliation{Helmholtz-Zentrum Dresden-Rossendorf, 01328 Dresden, Germany}

\author{Eric R. J. Edwards}
\affiliation{\mbox{Quantum Electromagnetics Division, National Institute of Standards and Technology, Boulder, CO 80305, USA}}

\author{Hans T. Nembach}
\affiliation{\mbox{Quantum Electromagnetics Division, National Institute of Standards and Technology, Boulder, CO 80305, USA}}

\author{Justin M. Shaw}
\affiliation{\mbox{Quantum Electromagnetics Division, National Institute of Standards and Technology, Boulder, CO 80305, USA}}

\author{Rudolf Gross}
\affiliation{\mbox{Walther-Meißner Institute, Bayerische Akademie der Wissenschaften, 85748 Garching, Germany}}
\affiliation{Physics Department, Technical University of Munich, 85748 Garching, Germany}
\affiliation{Nanosystems Initiative Munich, 80799 Munich, Germany}
\affiliation{\mbox{Munich Center for Quantum Science and Technology (MCQST), 80799 Munich, Germany}}

\author{Mathias Weiler}
\email{mathias.weiler@wmi.badw.de}
\affiliation{\mbox{Walther-Meißner Institute, Bayerische Akademie der Wissenschaften, 85748 Garching, Germany}}
\affiliation{Physics Department, Technical University of Munich, 85748 Garching, Germany}


\begin{abstract}
	We report ultra-low intrinsic magnetic damping in \CF \space heterostructures, reaching the low $10^{-4}$ regime at room temperature. By using a broadband ferromagnetic resonance technique in out-of-plane geometry, we extracted the dynamic magnetic properties of several \CF-based heterostructures with varying ferromagnetic layer thickness. By measuring radiative damping and spin pumping effects, we found the intrinsic damping of a 26\,nm thick sample to be $\alpha_{\mathrm{0}} \lesssim 3.18\times10^{-4}$. Furthermore, using Brillouin light scattering microscopy we measured spin-wave propagation lengths of up to $(21\pm1)\,\mathrm{\mu m}$ in a 26\,nm thick \CF \space heterostructure at room temperature, which is in excellent agreement with the measured damping.
\end{abstract}


\maketitle


%
%
%
%

Itinerant ferromagnets (FM) are advantageous for spintronic and magnonic devices. They benefit from, e.g., large magnetoresistive effects and current-induced spin-orbit torques~\cite{Gambardella2011a}. In many magneto-resistive technologies (e.g., anisotropic magnetoresistance, giant magnetoresistance, tunnel magnetoresistance) electronic conductivity is indispensable. Moreover, due to high saturation magnetization in metallic FMs, spin-wave (SW) group velocities are in general significantly higher than in insulating ferrimagnets~\cite{Wessels2016, Talalaevskij2017, Korner2017,Collet2017}. High saturation magnetizations in general ease detection. Nevertheless, itinerant FMs typically have considerable magnetic damping~\cite{Maksymov2015,Twisselmann2003}. This is unfavorable for many applications. For example, low damping is crucial for oscillators based on spin transfer torques and spin orbit torques as well as for achieving large spin-wave propagation lengths (SWPL)~\cite{Demidov2015,Kruglyak2010,Chumak2015}. The need for thin film materials with low magnetic damping has triggered the interest in the insulating ferrimagnet yttrium-iron garnet (Y$_{\text{3}}$Fe$_{\text{5}}$O$_{\text{12}}$, YIG)\cite{Hauser2016, Evelt2016, Jungfleisch2015}. Although for YIG, very small total (Gilbert) damping parameters in the order of $\alpha_\mathrm{G} \approx 10^{-5}$, and large SWPLs of a few tens of micrometers (up to $\sim$\,25\,$\mu $m) in thin films ($\sim$\,20\,nm) have been reported~\cite{Chang2014,Collet2017,Jungfleisch2015}, its insulating properties and requirement for crystalline growth are challenges for large scale magnonic applications.

Schoen \textit{et al.} recently observed ultra-low intrinsic magnetic damping in \CF \space (CoFe) metallic thin films ($\alpha_{\mathrm{0}} = (5 \pm 1.8)\times 10^{-4}$)\,\cite{Schoen2015}, and Körner \textit{et al.} reported PLs of $5\,\mathrm{\mu m}-8\,\mathrm{\mu m}$ in CoFe using time resolved scanning magneto-optical Kerr microscopy~\cite{Korner2017}. This motivated our study on sputter-deposited CoFe-based thin film heterostructures. We use broadband ferromagnetic resonance (BB-FMR) spectroscopy~\cite{Kalarickal2006} in out-of-plane (OOP) geometry and Brillouin light scattering (BLS) microscopy~\cite{Sebastian2015} and find intrinsic damping parameters in the lower $10^{-4}$ regime as well as SWPLs of more than $20\,\mathrm{\mu}$m. The damping is therefore comparable to YIG/heavy metal (HM) heterostructures~\cite{Sun2013} and the SWPL is comparable to that of state-of-the-art YIG thin films~\cite{Collet2017, Jungfleisch2015}. Thin film CoFe is a promising candidate for all-metal magnonic devices, as it combines low magnetic damping with good electrical conductivity and large saturation magnetization, while enabling easy fabrication by room-temperature processing/deposition, no required annealing, polycrystalline structure, and scalability to the nanometer regime.



For BB-FMR, Ta(3\,nm)/Al(3\,nm)/\CF($t$)/ Al(3\,nm)/Ta(3\,nm) heterostructures with different thickness $t$ of the CoFe layer were sputter deposited on a thermally oxidized Si (100) substrate at an Ar pressure of $5\times10^{-6}$\,bar at room temperature. No subsequent annealing process was performed. The CoFe layer thickness was varied between 1.4\,nm $< t <$ 26\,nm as determined by X-ray reflectometry. 



\begin{figure}[h]
	\centering
	\includegraphics{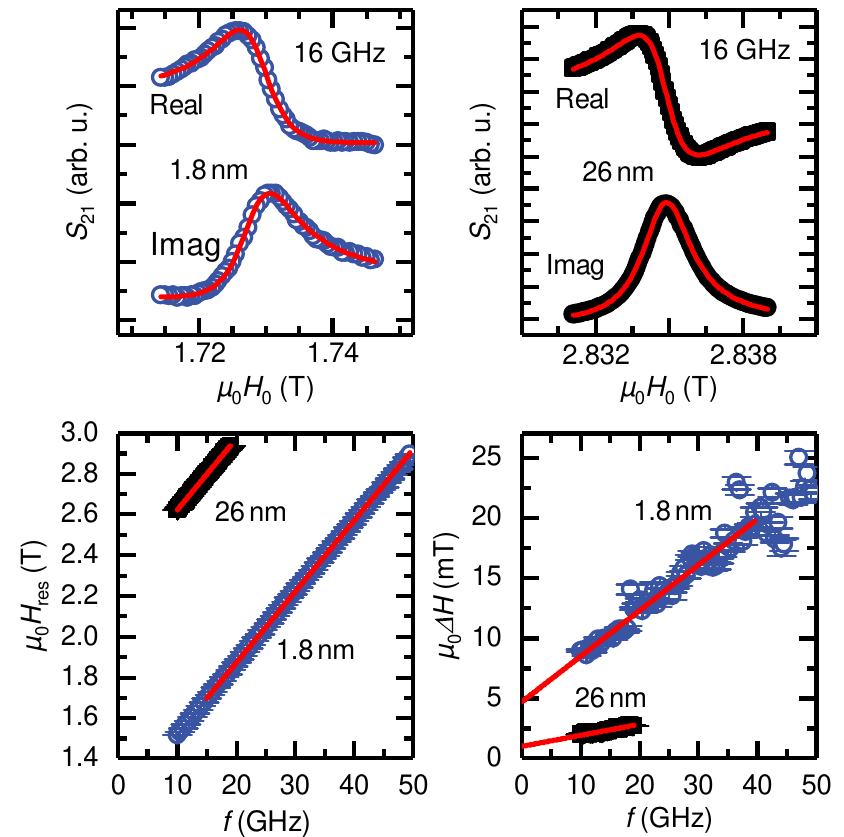}
	\caption{(a) Measured microwave transmission $S_\mathrm{21}$ at 16\,GHz vs. applied OOP magnetic field $H_\mathrm{0}$ for blanket Ta(3\,nm)/Al(3\,nm)/\CF($t$)/ Al(3\,nm)/Ta(3\,nm) samples with CoFe thickness $t=1.8$\,nm ((a) blue symbols) and $t=26$\,nm ((b) black symbols), respectively. The red lines are fits of Eq.\,\eqref{S21} to the data. The extracted resonance fields $H_\mathrm{res}$ and linewidths $\Delta H$ as a function of the applied microwave frequency are shown in (c) and (d), respectively. Here, the error bars (smaller than symbol size) are extracted fit errors from (a) and (b). In (c) the red line is a fit to Eq.~\eqref{OOP-Kittel} to extract the Landé-factor $g$ and the effective magnetization $M_{\mathrm{eff}}$. In (d), the linewidth is plotted vs. frequency. The Gilbert parameter $\alpha_{\mathrm{G}}$ and the inhomogeneous linewidth broadening $H_{\mathrm{inh}}$ are extracted by fitting the data to Eq.~\eqref{Alpha-fit} (red lines). The linewidth of the $t=26\,\mathrm{nm}$ thick sample is shown in Fig.\,\ref{Panel2c}\,(c) on an expanded scale.}
	\label{Panel1}
\end{figure}

The OOP BB-FMR measurements were performed at room temperature with a vector network analyzer (VNA). This geometry was chosen to determine the intrinsic magnetic damping without further damping contributions due to magnon-magnon scattering~\cite{Hillebrands2003}. The samples were placed directly on a coplanar waveguide (CPW), with a $80\,\mathrm{\mu m} $ wide center conductor. For the measurements, the VNA frequency $f$ was kept constant and the microwave transmission parameter $S_\mathrm{21}$ was recorded as a function of applied magnetic field $H_\mathrm{0}$ for a range of frequencies at a VNA output power of 0\,dBm.
A representative set of data as measured of the real and imaginary part of $S_\mathrm{21}$ at 16\,GHz for samples with $t = 1.8\,\mathrm{nm}$ and $t = 26\,\mathrm{nm}$ is shown in Fig.\,\ref{Panel1}\,(a) and (b).

The magnetic response of the thin film FM magnetized out-of-plane is given by the susceptibility $\chi$
which is obtained by solving the Landau-Lifshitz-Gilbert (LLG) equation\,\cite{Nembach2011, Schoen2015}:
\begin{eqnarray}
\chi (H_\mathrm{0}) &=& \frac{M_\mathrm{s} (H_\mathrm{0} - H_\mathrm{res} + H_\mathrm{eff})}{(H_\mathrm{0} - H_\mathrm{res} + H_\mathrm{eff} + i \frac{\Delta H}{2})^2 - H_\mathrm{eff}^2}.
\end{eqnarray}
Here, $M_\mathrm{s}$ is the saturation magnetization, $H_\mathrm{res}$ is the resonance field, $H_\mathrm{eff} = 2 \pi f /(\mu_\mathrm{0} \gamma)$ with $\gamma$ being the gyromagnetic ratio and $\Delta H = 2(2 \pi f \alpha)/(\gamma \mu_{\mathrm{0}}) $ is the full width at half maximum (FWHM) linewidth of the resonance.
The data in Fig\,\ref{Panel1}\,(a) and (b) is fitted to\,\cite{Berger2018}
\begin{equation}\label{S21}
S_\mathrm{21}(H_\mathrm{0}) = S_\mathrm{21}^{0} + i A \frac{\chi (H_\mathrm{0})}{M_\mathrm{s}} = S_\mathrm{21}^{0} (1 + \Delta S_\mathrm{21}),
\end{equation}
where $S_\mathrm{21}^{0}$ is the background transmission through the CPW without magnetic resonance peak. It is determined from the fits as a complex linear background to the data $S_\mathrm{21}^{0}(H_\mathrm{0}) = S_\mathrm{21}^{a} + H_\mathrm{0} S_\mathrm{21}^{b}$. The factor $A$ is a complex-valued scaling parameter.

In the OOP geometry, the resonance condition for thin films is given by~\cite{Kittel1948}
\begin{equation}\label{OOP-Kittel}
\mu_{\mathrm{0}}\,H_\text{res} =\mu_{\mathrm{0}}\, M_\text{eff} + \mu_{\mathrm{0}}\, H_\text{eff},
\end{equation}
where $M_\mathrm{eff} = M_\mathrm{s} - H_\mathrm{k} $ is the effective magnetization, with the uniaxial out-of-plane anisotropy field $H_\mathrm{k}$. In Fig.\,\ref{Panel1}\,(c), we plot the determined $H_\mathrm{res}$ vs. the frequency $f$. From the fit to Eq.\,\eqref{OOP-Kittel} (red solid lines in Fig.\,\ref{Panel1}\,(c)), we obtain $M_\mathrm{eff}$ and $\gamma$ of the specific sample.

The FWHM linewidth vs. frequency data shown in Fig.\,\ref{Panel1}\,(d) is fitted to 
\begin{equation}\label{Alpha-fit}
\mu_{\mathrm{0}}\,\Delta H = \mu_{\mathrm{0}}\,H_{\mathrm{inh}} + 2 \cdot \frac{2\pi f \alpha_\mathrm{G}}{\gamma}.
\end{equation}
Here, $H_\mathrm{inh}$ is the inhomogeneous linewidth broadening and $\alpha_\mathrm{G}$ is the phenomenological Gilbert damping parameter~\cite{Woltersdorf2013,McMichael2003}. $H_\mathrm{inh}$ indicates the presence of long-range magnetic inhomogeneities, which become more relevant for thinner films, but do not contribute to our $\alpha_\mathrm{G}$.

\begin{figure}[!]
	\centering
	\includegraphics{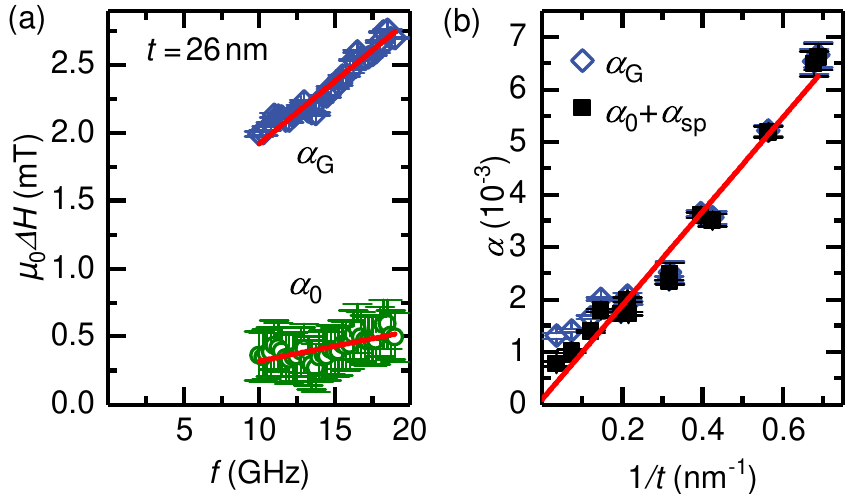}
	\caption{(a) An expanded view of the linewidth vs. frequency plot of the $t=26\,\mathrm{nm}$ sample. The total linewidth is shown by the blue diamonds, from which the total Gilbert damping parameter $\alpha_{\mathrm{G}}$ was extracted. The green circles represent the intrinsic linewidth contribution. In (b), the total damping $\alpha_{\mathrm{G}}$ is plotted for different thicknesses $t$ as blue diamonds. We substracted the contributions from radiative damping and eddy currents and show the resulting $\alpha_{\mathrm{0}} + \alpha_{\mathrm{sp}}$ as black squares. The red line is an unweighted fit to Eq.\,\eqref{AlphaSP} in order to quantify the spin pumping contribution within our samples and to be able to extrapolate the intrinsic damping of CoFe within our multilayer system. For thicker samples, the available frequency range is rather small, leading to an increased uncertainty, as discussed in Ref.\citenum{Shaw2013a}.}
	\label{Panel2c}
\end{figure}

Several contributions to the measured total damping ($\alpha_{\mathrm{G}}$) were extracted from our data. In addition to the intrinsic damping of the magnetic material itself ($\alpha_{\mathrm{0}}$), spin pumping ($\alpha_{\mathrm{sp}}$) contributes significantly~\cite{Tserkovnyak2002,Haertinger2015,Brataas2017} to the total damping in our thinner heterostructures due to the adjacent HM (Ta) layers. Furthermore, we consider additional damping contributions from eddy currents ($\alpha_{\mathrm{eddy}}$) and radiative damping ($\alpha_{\mathrm{rad}}$)\cite{Schoen2015,Berger2018}. Due to these contributions, the total damping ($\alpha_{\mathrm{G}} = \alpha_{\mathrm{0}} + \alpha_{\mathrm{sp}} + \alpha_{\mathrm{eddy}} + \alpha_{\mathrm{rad}}$) depends on the FM thickness. We calculated damping due to eddy currents and measured radiative damping contributions to the total damping. The eddy current contribution is given by\cite{Schoen2015} $\alpha_\mathrm{eddy} = \gamma \mu_{\mathrm{0}}^2 M_\mathrm{s}t^2/16\rho $. Here, $\mu_{\mathrm{0}}M_\mathrm{s}= 2.35$\,T (see SI) and $\rho = 340\,\mathrm{n\Omega\,m}$  is the estimated weighted resistivity value of the CoFe film derived from the resistivities of iron and cobalt thin films with thicknesses of around 20\,nm\cite{Raeburn1978, DeVries1988}. With these values, we find an almost negligible eddy current contribution to the total damping.
A quantitative determination analogous to Ref.\,\citenum{Berger2018} of the radiative damping is done by analyzing the magnitude of the measured inductance $L$ of all samples. The quantification of this contribution is important for BB-FMR, because it represents a damping by inductive power dissipation into the CPW and, hence, is not a property of the sample itself but depends on the setup. In possible applications like, e.g., magnonic waveguides or spin-Hall nano-oscillators, this contribution vanishes and the damping lowers by $\alpha_{\mathrm{rad}}$.
With Eq.\,\eqref{S21} above and Eq.\,(9) from Ref.\,\citenum{Berger2018}, one obtains:
\begin{equation}
\frac{L}{\chi} \equiv \tilde{L} = - \frac{2 Z_\mathrm{0} A}{M_\mathrm{s}S_\mathrm{21}^{0} \omega}.
\end{equation}
Here, $Z_\mathrm{0} = 50\,\mathrm{\Omega}$ is the CPW impedance. It has been shown, that $\tilde{L} = \tilde{L}_\mathrm{0} + \tilde{L}_\mathrm{1}(\omega),$ where $\tilde{L}_\mathrm{0} \in \mathbb{R}$ and $\tilde{L}_\mathrm{1} \in \mathbb{C}$, due to the effect of inverse spin-orbit torques~\cite{Berger2018}. We extract $L$ from the FMR measurements, and the dipolar inductance $\tilde{L}_\mathrm{0}$ from a fit of $\tilde{L}$ vs. $f$ for each sample. The radiative damping contribution is then given as\cite{Schoen2015}
\begin{equation}
\alpha_{\mathrm{rad}} = \frac{1}{4} \frac{\gamma \mu_{\mathrm{0}} M_\mathrm{s}}{Z_\mathrm{0}} \tilde{L}_\mathrm{0}.
\end{equation}
This analysis allows us to determine $\alpha_{\mathrm{rad}}$ independently of geometrical parameters of the samples or CPWs and is used to quantitatively extract the dipolar inductance without any calibration of the microwave circuit. For the thickest sample we obtain $\alpha_{\mathrm{rad}} = (4.69 \pm 0.05) \times 10^{-4}$, which is comparable to previously obtained values~\cite{Schoen2015, Schoen2016}. 
The damping including the spin pumping contribution $\alpha_{\mathrm{sp}}$ is given by
\begin{equation}\label{AlphaSP}
\alpha_\mathrm{0} + \alpha_{\mathrm{sp}} =\alpha_\mathrm{0} + 2 \frac{\gamma \hbar g_\mathrm{eff}^{\uparrow \downarrow}}{4 \pi M_\text{s}} \frac{1}{t},
\end{equation}
where $g_\mathrm{eff}^{\uparrow \downarrow}$ is the effective spin mixing conductance\cite{Schoen2016}.  We substract $\alpha_{\mathrm{rad}}$ and $\alpha_{\mathrm{eddy}}$ from the measured total damping $\alpha_{\mathrm{G}}$ (see Fig.\,\ref{Panel2c}\,(a) and (b)) and plot the remaining damping $\alpha_{\mathrm{0}} + \alpha_{\mathrm{sp}}$ as a function of 1/$t$ in Fig.\,\ref{Panel2c}\,(b) together with the total damping $\alpha_{\mathrm{G}}$. From a linear fit (Eq.\,\eqref{AlphaSP}) to $\alpha_{\mathrm{0}} + \alpha_{\mathrm{sp}}$, we obtain $g_\mathrm{eff}$ and $\alpha_{\mathrm{0}}$. Herefore, we use $M_\mathrm{s}$ as above and $\gamma/2 \pi = 28.65\, \mathrm{GHz/T}$. The fitted $g_\mathrm{eff} = (5.5 \pm 0.3) \times 10^{18}\,\mathrm{m^{-2}}$ is in agreement with literature values\cite{Schoen2015}. The $y$-intercept indicating the extrapolated intrinsic damping yields $\alpha_{\mathrm{0}} = (0.91 \pm 1.69)\times 10^{-4}$ hence, the intrinsic damping is below the sensitivity of our approach. For the thickest sample $t=26\,\mathrm{nm}$ shown in Fig.\,\ref{Panel2c}\,(a), we obtain $\alpha_{\mathrm{0}} = (3.18 \pm 0.48) \times 10^{-4}$ (see SI for details). Within the errors, this value lies close to the extrapolated value and is the lowest intrinsic damping for a thin film ferromagnetic metal reported so far. We attribute the slightly reduced intrinsic $\alpha_\mathrm{0}$ compared to Ref.\,\citenum{Schoen2016} to the use of a different seed layer, which has a substantial impact on the damping of CoFe\cite{Edwards2019}.

\begin{figure*}[t]
	\centering
	\includegraphics{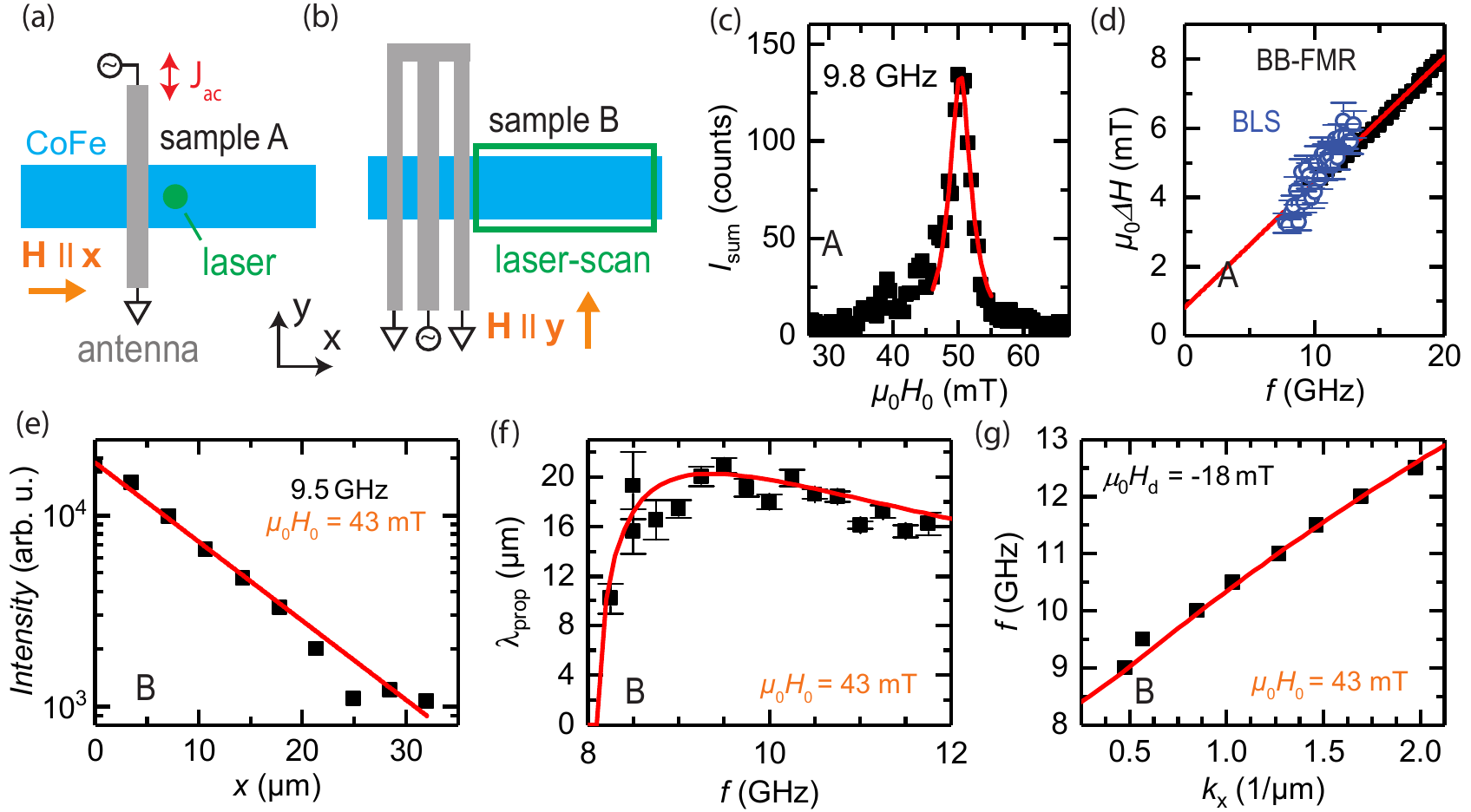}
	\caption{(a), (b) Schematic top view of sample A and B, respectively. (c) Integrated BLS intensity vs. field of sample A. By use of a Lorentzian fit (red line), the linewidth is extracted. In (d) we compare BLS linewidth (open symbols) to values obtained by in-plane BB-FMR on a blanket film (closed squares). (e) Respresentative data set of sample B with $f=9.5$\,GHz. The BLS intensity was measured as a function of the position ($x$,$y$) in the area highlighted with the green rectangle in (b). The measured signal was then integrated in $y$ direction and the exponential decay in $x$ direction is fitted (red curve). (f) Propagation length $\lambda_\mathrm{prop}$ for varying frequency $f$. Depicted error bars are fit errors. The red curve is based on an analytical model calculation (see text). (g) $f$ vs. $k_\mathrm{x}$ dispersion determined by phase-resolved $\mu$BLS. The red line is a model from Eq.\,\eqref{KalinikosSlavin}.}
	\label{Panel3f}
\end{figure*}

The low damping properties of the CoFe heterostructures, in combination with the high saturation magnetization are expected to result in long PLs of dipolar SWs. We use microfocused BLS\cite{Sebastian2015} to study the SW propagation in patterned CoFe samples, which are schematically depicted in Fig.\,\ref{Panel3f}\,(a) and (b).

For our experiments, we fabricated patterned stripes of a Pt(3\,nm)/Cu(3\,nm)/\CF($t$)/Cu(3\,nm)/Ta(3\,nm) heterostructure using laser (sample A) and electron beam (sample B) lithography, sputter deposition and a subsequent lift-off process. This stack sequence was used as lower in-plane damping was observed compared to the samples containing Al. Below, we present data on only two samples with a thickness of $t = 5\,\mathrm{nm}$ and a width $w = 1.5\,\mathrm{\mu m}$ for sample A and $t = 26\,\mathrm{nm} $ and  $w = 5\,\mathrm{\mu m}$ for sample B, respectively. An aluminum antenna was placed on top of the CoFe strip to drive spin dynamics via a microwave drive applied to the antenna. For sample A we used a simple aluminum strip optimized for excitation of the uniform (FMR) mode, whereas for the sample B we used a CPW antenna optimized for an efficient excitation of SWs with wave number $ k \leq 2\,\mathrm{\mu m^{-1}}$.

In order to compare the uniform FMR-mode linewidths of extended and patterned films, we used sample A in backward volume geometry and placed the laser spot close to the antenna, where the FMR mode is dominantly excited. We recorded BLS spectra for several magnetic fields for each frequency. The BLS intensity is integrated and the signal sum is then plotted vs. the external magnetic field in Fig.\,\ref{Panel3f}\,(c). The FWHM-linewidth $\Delta H$ is determined by fitting a Lorentzian (red line). We then compared the fitted linewidth to the measured in-plane BB-FMR linewidth of a blanket film, deposited simultaneously with the structured BLS sample. In the in-plane configuration the total damping increases due to magnon-magnon scattering~\cite{Hillebrands2003, Arias1999} and possible anisotropic damping~\cite{Chen2018,Seib2009,Steiauf2005,Safonov2002}. As shown in Fig.\,\ref{Panel3f}\,(d), the linewidths $\mu_{\mathrm{0}}\Delta H$ determined from BB-FMR (black sybmbols) and BLS (blue symbols) are very similar, indicating that the damping properties are not affected by the patterning, as expected in a lift-off process with micrometer feature sizes.

In the next set of experiments, we investigate the SWPL of sample B (see Fig.\,\ref{Panel3f}\,(b)). Here, the magnitude of the external magnetic field was fixed at $\mu_{\mathrm{0}}H_\mathrm{0} = 43$\,mT, while the field was applied perpendicular to the CoFe strip (Damon-Eshbach geometry). The BLS intensity was recorded as a function of position ($x$,$y$) over the CoFe strip. The BLS intensity decay in $x$ direction (i.e. the BLS intensity averaged over the width of the strip in order to suppress mode-beating effects~\cite{Pirro2011,Demidov2009,Clausen2011}) is shown in Fig.\,\ref{Panel3f}\,(e) for $f = 9.5$\,GHz. The SWPL $\lambda_\mathrm{prop}$ is extracted by a fit to $I = I_\mathrm{0} \exp(-2x/\lambda_\mathrm{prop})$~\cite{Demidov2016} and plotted vs. $f$ in Fig.\,\ref{Panel3f}\,(f). From our experiments, we extract a maximum SWPL of $(21\pm1)\,\mathrm{\mu m}$, well exceeding previously obtained results for FeNi alloys~\cite{Yamanoi2013} and CoFe\,\cite{Korner2017} and very comparable to values found for YIG thin films~\cite{Collet2017,Jungfleisch2015}. The red curve is the theoretical prediction, based on the analytical Kalinikos-Slavin model detailed below and using the magnetic parameters determined by in-plane BB-FMR ($\mu_{\mathrm{0}}M_\mathrm{s} = 2.35$\,T, $\mu_{\mathrm{0}}M_\mathrm{eff} = 2.29$\,T, $\alpha_{\mathrm{G}} - \alpha_{\mathrm{rad}} = 3.92\times10^{-3}$, $g=2.051$) for a co-deposited reference sample (see SI).

Starting with a simplified version of Kalinikos and Slavin's SW dispersion for the modes with $\mathbf{k_\mathrm{x}} \perp \mathbf{M}$\cite{Kalinikos1986, Liensberger2019}
\begin{multline}\label{KalinikosSlavin}
f_\mathrm{res} = \frac{\mu_{\mathrm{0}} \gamma}{2 \pi} \sqrt{H_\mathrm{0} + H_\mathrm{d} + H_\mathrm{k} + M_\mathrm{s}\frac{1-\exp(-k t)}{k t}} \\
\times \sqrt{H_\mathrm{0} + H_\mathrm{d} + M_\mathrm{s} \left(1 - \frac{1-\exp(-k t)}{k t}\right)},
\end{multline} 
we calculated the group velocity $v_\mathrm{g} = 2\pi \partial f_\mathrm{res}/\partial k$. Here, $k = \sqrt{k_\mathrm{x}^2 + k_\mathrm{y}^2}$ is the in-plane wave vector of the travelling SW and $\mu_{\mathrm{0}}H_\mathrm{k} = \mu_{\mathrm{0}}M_\mathrm{eff} - \mu_{\mathrm{0}}M_\mathrm{s} = -60\,\mathrm{mT}$ is the effective interface anisotropy field. The calculation of the transversal wave vector component $k_\mathrm{y} = 0.31\,\mathrm{\mu m^{-1}}$ due to geometrical confinement was shown to be non-trivial and is used as a fitting parameter, as in Ref.\,\citenum{Duan2015}. The resonance linewidth is given by\cite{Stancil2009} $\Delta \omega = \alpha \mu_{\mathrm{0}} \gamma (M_\mathrm{eff}/2 + H_\mathrm{0} + H_\mathrm{d})$ and the lifetime of the SW is $\tau = 1/\Delta \omega$. Here, $\alpha = \alpha_{\mathrm{G}} - \alpha_{\mathrm{rad}}$. The SWPL is $\lambda_\mathrm{prop} = v_\mathrm{g} \tau$. The demagnetization field in $y$-direction was set to $\mu_{\mathrm{0}}H_\mathrm{d} = -18$\,mT, as required for matching Eq.\,\eqref{KalinikosSlavin} to the SW dispersion obtained by phase-resolved $\mu$BLS\,\cite{Sebastian2015} (see Fig.\,\ref{Panel3f}\,(g)). This value for $H_\mathrm{d}$ is in good agreement with the demagnetization
($\mu_{\mathrm{0}}H_\mathrm{d} \approx -12$\,mT) obtained for an ellipsoid with the axes corresponding to the CoFe-stripe dimensions\,\cite{Osborn1945}. We find excellent agreement between this model and our experimental data in Fig.\,\ref{Panel3f}\,(f).


In summary, our sputter-deposited \CF \space layers exhibit a record low intrinsic damping for metallic thin film ferromagnets of $\alpha_{\mathrm{0}} \lesssim 3.18\times 10^{-4}$ in OOP geometry. The damping properties of extended films are maintained for micropatterned films, and spin-wave propagation lengths are in very good agreement with the properties extracted from BB-FMR. The low magnetic damping, together with the high saturation magnetization, lead to spin-wave decay lengths of more than 20\,$\mathrm{\mu m}$ at room temperature, which are the highest reported so far in itinerant magnetic systems. This property makes \CF \space a promising material for all-metal spintronic and magnonic devices, compatible with semiconductor technology.
\begin{acknowledgements}
	We acknowledge financial support by the Deutsche Forschungsgemeinschaft (DFG, German Research Foundation) via WE5386/4, WE5386/5 and Germany's Excellence Strategy EXC-2111-390814868.
\end{acknowledgements}


\begin{thebibliography}{47}%
	\makeatletter
	\providecommand \@ifxundefined [1]{%
		\@ifx{#1\undefined}
	}%
	\providecommand \@ifnum [1]{%
		\ifnum #1\expandafter \@firstoftwo
		\else \expandafter \@secondoftwo
		\fi
	}%
	\providecommand \@ifx [1]{%
		\ifx #1\expandafter \@firstoftwo
		\else \expandafter \@secondoftwo
		\fi
	}%
	\providecommand \natexlab [1]{#1}%
	\providecommand \enquote  [1]{``#1''}%
	\providecommand \bibnamefont  [1]{#1}%
	\providecommand \bibfnamefont [1]{#1}%
	\providecommand \citenamefont [1]{#1}%
	\providecommand \href@noop [0]{\@secondoftwo}%
	\providecommand \href [0]{\begingroup \@sanitize@url \@href}%
	\providecommand \@href[1]{\@@startlink{#1}\@@href}%
	\providecommand \@@href[1]{\endgroup#1\@@endlink}%
	\providecommand \@sanitize@url [0]{\catcode `\\12\catcode `\$12\catcode
		`\&12\catcode `\#12\catcode `\^12\catcode `\_12\catcode `\%12\relax}%
	\providecommand \@@startlink[1]{}%
	\providecommand \@@endlink[0]{}%
	\providecommand \url  [0]{\begingroup\@sanitize@url \@url }%
	\providecommand \@url [1]{\endgroup\@href {#1}{\urlprefix }}%
	\providecommand \urlprefix  [0]{URL }%
	\providecommand \Eprint [0]{\href }%
	\providecommand \doibase [0]{http://dx.doi.org/}%
	\providecommand \selectlanguage [0]{\@gobble}%
	\providecommand \bibinfo  [0]{\@secondoftwo}%
	\providecommand \bibfield  [0]{\@secondoftwo}%
	\providecommand \translation [1]{[#1]}%
	\providecommand \BibitemOpen [0]{}%
	\providecommand \bibitemStop [0]{}%
	\providecommand \bibitemNoStop [0]{.\EOS\space}%
	\providecommand \EOS [0]{\spacefactor3000\relax}%
	\providecommand \BibitemShut  [1]{\csname bibitem#1\endcsname}%
	\let\auto@bib@innerbib\@empty
	\bibitem [{\citenamefont {Gambardella}\ and\ \citenamefont
		{Miron}(2011)}]{Gambardella2011a}%
	\BibitemOpen
	\bibfield  {author} {\bibinfo {author} {\bibfnamefont {P.}~\bibnamefont
			{Gambardella}}\ and\ \bibinfo {author} {\bibfnamefont {I.~M.}\ \bibnamefont
			{Miron}},\ }\href {\doibase 10.1098/rsta.2010.0336} {\bibfield  {journal}
		{\bibinfo  {journal} {Philosophical Transactions of the Royal Society A:
				Mathematical, Physical and Engineering Sciences}\ }\textbf {\bibinfo {volume}
			{369}},\ \bibinfo {pages} {3175} (\bibinfo {year} {2011})}\BibitemShut
	{NoStop}%
	\bibitem [{\citenamefont {Wessels}\ \emph {et~al.}(2016)\citenamefont
		{Wessels}, \citenamefont {Vogel}, \citenamefont {T{\"{o}}dt}, \citenamefont
		{Wieland}, \citenamefont {Meier},\ and\ \citenamefont
		{Drescher}}]{Wessels2016}%
	\BibitemOpen
	\bibfield  {author} {\bibinfo {author} {\bibfnamefont {P.}~\bibnamefont
			{Wessels}}, \bibinfo {author} {\bibfnamefont {A.}~\bibnamefont {Vogel}},
		\bibinfo {author} {\bibfnamefont {J.~N.}\ \bibnamefont {T{\"{o}}dt}},
		\bibinfo {author} {\bibfnamefont {M.}~\bibnamefont {Wieland}}, \bibinfo
		{author} {\bibfnamefont {G.}~\bibnamefont {Meier}}, \ and\ \bibinfo {author}
		{\bibfnamefont {M.}~\bibnamefont {Drescher}},\ }\href {\doibase
		10.1038/srep22117} {\bibfield  {journal} {\bibinfo  {journal} {Scientific
				Reports}\ }\textbf {\bibinfo {volume} {6}},\ \bibinfo {pages} {1} (\bibinfo
		{year} {2016})}\BibitemShut {NoStop}%
	\bibitem [{\citenamefont {Talalaevskij}\ \emph {et~al.}(2017)\citenamefont
		{Talalaevskij}, \citenamefont {Decker}, \citenamefont {Stigloher},
		\citenamefont {Mitra}, \citenamefont {K{\"{o}}rner}, \citenamefont
		{Cespedes}, \citenamefont {Back},\ and\ \citenamefont
		{Hickey}}]{Talalaevskij2017}%
	\BibitemOpen
	\bibfield  {author} {\bibinfo {author} {\bibfnamefont {A.}~\bibnamefont
			{Talalaevskij}}, \bibinfo {author} {\bibfnamefont {M.}~\bibnamefont
			{Decker}}, \bibinfo {author} {\bibfnamefont {J.}~\bibnamefont {Stigloher}},
		\bibinfo {author} {\bibfnamefont {A.}~\bibnamefont {Mitra}}, \bibinfo
		{author} {\bibfnamefont {H.~S.}\ \bibnamefont {K{\"{o}}rner}}, \bibinfo
		{author} {\bibfnamefont {O.}~\bibnamefont {Cespedes}}, \bibinfo {author}
		{\bibfnamefont {C.~H.}\ \bibnamefont {Back}}, \ and\ \bibinfo {author}
		{\bibfnamefont {B.~J.}\ \bibnamefont {Hickey}},\ }\href {\doibase
		10.1103/PhysRevB.95.064409} {\bibfield  {journal} {\bibinfo  {journal}
			{Physical Review B}\ }\textbf {\bibinfo {volume} {95}},\ \bibinfo {pages} {1}
		(\bibinfo {year} {2017})}\BibitemShut {NoStop}%
	\bibitem [{\citenamefont {K{\"{o}}rner}\ \emph {et~al.}(2017)\citenamefont
		{K{\"{o}}rner}, \citenamefont {Schoen}, \citenamefont {Mayer}, \citenamefont
		{Decker}, \citenamefont {Stigloher}, \citenamefont {Weindler}, \citenamefont
		{Meier}, \citenamefont {Kronseder},\ and\ \citenamefont {Back}}]{Korner2017}%
	\BibitemOpen
	\bibfield  {author} {\bibinfo {author} {\bibfnamefont {H.~S.}\ \bibnamefont
			{K{\"{o}}rner}}, \bibinfo {author} {\bibfnamefont {M.~A.~W.}\ \bibnamefont
			{Schoen}}, \bibinfo {author} {\bibfnamefont {T.}~\bibnamefont {Mayer}},
		\bibinfo {author} {\bibfnamefont {M.~M.}\ \bibnamefont {Decker}}, \bibinfo
		{author} {\bibfnamefont {J.}~\bibnamefont {Stigloher}}, \bibinfo {author}
		{\bibfnamefont {T.}~\bibnamefont {Weindler}}, \bibinfo {author}
		{\bibfnamefont {T.~N.~G.}\ \bibnamefont {Meier}}, \bibinfo {author}
		{\bibfnamefont {M.}~\bibnamefont {Kronseder}}, \ and\ \bibinfo {author}
		{\bibfnamefont {C.~H.}\ \bibnamefont {Back}},\ }\href {\doibase
		10.1063/1.4994137} {\bibfield  {journal} {\bibinfo  {journal} {Applied
				Physics Letters}\ }\textbf {\bibinfo {volume} {111}},\ \bibinfo {pages}
		{132406} (\bibinfo {year} {2017})}\BibitemShut {NoStop}%
	\bibitem [{\citenamefont {Collet}\ \emph {et~al.}(2017)\citenamefont {Collet},
		\citenamefont {Gladii}, \citenamefont {Evelt}, \citenamefont {Bessonov},
		\citenamefont {Soumah}, \citenamefont {Bortolotti}, \citenamefont
		{Demokritov}, \citenamefont {Henry}, \citenamefont {Cros}, \citenamefont
		{Bailleul}, \citenamefont {Demidov},\ and\ \citenamefont
		{Anane}}]{Collet2017}%
	\BibitemOpen
	\bibfield  {author} {\bibinfo {author} {\bibfnamefont {M.}~\bibnamefont
			{Collet}}, \bibinfo {author} {\bibfnamefont {O.}~\bibnamefont {Gladii}},
		\bibinfo {author} {\bibfnamefont {M.}~\bibnamefont {Evelt}}, \bibinfo
		{author} {\bibfnamefont {V.}~\bibnamefont {Bessonov}}, \bibinfo {author}
		{\bibfnamefont {L.}~\bibnamefont {Soumah}}, \bibinfo {author} {\bibfnamefont
			{P.}~\bibnamefont {Bortolotti}}, \bibinfo {author} {\bibfnamefont {S.~O.}\
			\bibnamefont {Demokritov}}, \bibinfo {author} {\bibfnamefont
			{Y.}~\bibnamefont {Henry}}, \bibinfo {author} {\bibfnamefont
			{V.}~\bibnamefont {Cros}}, \bibinfo {author} {\bibfnamefont {M.}~\bibnamefont
			{Bailleul}}, \bibinfo {author} {\bibfnamefont {V.~E.}\ \bibnamefont
			{Demidov}}, \ and\ \bibinfo {author} {\bibfnamefont {A.}~\bibnamefont
			{Anane}},\ }\href {\doibase 10.1063/1.4976708} {\bibfield  {journal}
		{\bibinfo  {journal} {Applied Physics Letters}\ }\textbf {\bibinfo {volume}
			{110}},\ \bibinfo {pages} {092408} (\bibinfo {year} {2017})}\BibitemShut
	{NoStop}%
	\bibitem [{\citenamefont {Maksymov}\ and\ \citenamefont
		{Kostylev}(2015)}]{Maksymov2015}%
	\BibitemOpen
	\bibfield  {author} {\bibinfo {author} {\bibfnamefont {I.~S.}\ \bibnamefont
			{Maksymov}}\ and\ \bibinfo {author} {\bibfnamefont {M.}~\bibnamefont
			{Kostylev}},\ }\href {\doibase 10.1016/j.physe.2014.12.027} {\bibfield
		{journal} {\bibinfo  {journal} {Physica E: Low-Dimensional Systems and
				Nanostructures}\ }\textbf {\bibinfo {volume} {69}},\ \bibinfo {pages} {253}
		(\bibinfo {year} {2015})}\BibitemShut {NoStop}%
	\bibitem [{\citenamefont {Twisselmann}\ and\ \citenamefont
		{McMichael}(2003)}]{Twisselmann2003}%
	\BibitemOpen
	\bibfield  {author} {\bibinfo {author} {\bibfnamefont {D.~J.}\ \bibnamefont
			{Twisselmann}}\ and\ \bibinfo {author} {\bibfnamefont {R.~D.}\ \bibnamefont
			{McMichael}},\ }\href {\doibase 10.1063/1.1543884} {\bibfield  {journal}
		{\bibinfo  {journal} {Journal of Applied Physics}\ }\textbf {\bibinfo
			{volume} {93}},\ \bibinfo {pages} {6903} (\bibinfo {year}
		{2003})}\BibitemShut {NoStop}%
	\bibitem [{\citenamefont {Demidov}\ \emph {et~al.}(2014)\citenamefont
		{Demidov}, \citenamefont {Urazhdin}, \citenamefont {Zholud}, \citenamefont
		{Sadovnikov},\ and\ \citenamefont {Demokritov}}]{Demidov2015}%
	\BibitemOpen
	\bibfield  {author} {\bibinfo {author} {\bibfnamefont {V.~E.}\ \bibnamefont
			{Demidov}}, \bibinfo {author} {\bibfnamefont {S.}~\bibnamefont {Urazhdin}},
		\bibinfo {author} {\bibfnamefont {A.}~\bibnamefont {Zholud}}, \bibinfo
		{author} {\bibfnamefont {A.~V.}\ \bibnamefont {Sadovnikov}}, \ and\ \bibinfo
		{author} {\bibfnamefont {S.~O.}\ \bibnamefont {Demokritov}},\ }\href
	{\doibase 10.1063/1.4901027} {\bibfield  {journal} {\bibinfo  {journal}
			{Applied Physics Letters}\ }\textbf {\bibinfo {volume} {105}},\ \bibinfo
		{pages} {172410} (\bibinfo {year} {2014})}\BibitemShut {NoStop}%
	\bibitem [{\citenamefont {Kruglyak}, \citenamefont {Demokritov},\ and\
		\citenamefont {Grundler}(2010)}]{Kruglyak2010}%
	\BibitemOpen
	\bibfield  {author} {\bibinfo {author} {\bibfnamefont {V.~V.}\ \bibnamefont
			{Kruglyak}}, \bibinfo {author} {\bibfnamefont {S.~O.}\ \bibnamefont
			{Demokritov}}, \ and\ \bibinfo {author} {\bibfnamefont {D.}~\bibnamefont
			{Grundler}},\ }\href {\doibase 10.1088/0022-3727/43/26/264001} {\bibfield
		{journal} {\bibinfo  {journal} {Journal of Physics D: Applied Physics}\
		}\textbf {\bibinfo {volume} {43}},\ \bibinfo {pages} {264001} (\bibinfo
		{year} {2010})}\BibitemShut {NoStop}%
	\bibitem [{\citenamefont {Chumak}\ \emph {et~al.}(2015)\citenamefont {Chumak},
		\citenamefont {Vasyuchka}, \citenamefont {Serga},\ and\ \citenamefont
		{Hillebrands}}]{Chumak2015}%
	\BibitemOpen
	\bibfield  {author} {\bibinfo {author} {\bibfnamefont {A.~V.}\ \bibnamefont
			{Chumak}}, \bibinfo {author} {\bibfnamefont {V.~I.}\ \bibnamefont
			{Vasyuchka}}, \bibinfo {author} {\bibfnamefont {A.~A.}\ \bibnamefont
			{Serga}}, \ and\ \bibinfo {author} {\bibfnamefont {B.}~\bibnamefont
			{Hillebrands}},\ }\href {\doibase 10.1038/nphys3347} {\bibfield  {journal}
		{\bibinfo  {journal} {Nature Physics}\ }\textbf {\bibinfo {volume} {11}},\
		\bibinfo {pages} {453} (\bibinfo {year} {2015})}\BibitemShut {NoStop}%
	\bibitem [{\citenamefont {Hauser}\ \emph {et~al.}(2016)\citenamefont {Hauser},
		\citenamefont {Richter}, \citenamefont {Homonnay}, \citenamefont
		{Eisenschmidt}, \citenamefont {Qaid}, \citenamefont {Deniz}, \citenamefont
		{Hesse}, \citenamefont {Sawicki}, \citenamefont {Ebbinghaus},\ and\
		\citenamefont {Schmidt}}]{Hauser2016}%
	\BibitemOpen
	\bibfield  {author} {\bibinfo {author} {\bibfnamefont {C.}~\bibnamefont
			{Hauser}}, \bibinfo {author} {\bibfnamefont {T.}~\bibnamefont {Richter}},
		\bibinfo {author} {\bibfnamefont {N.}~\bibnamefont {Homonnay}}, \bibinfo
		{author} {\bibfnamefont {C.}~\bibnamefont {Eisenschmidt}}, \bibinfo {author}
		{\bibfnamefont {M.}~\bibnamefont {Qaid}}, \bibinfo {author} {\bibfnamefont
			{H.}~\bibnamefont {Deniz}}, \bibinfo {author} {\bibfnamefont
			{D.}~\bibnamefont {Hesse}}, \bibinfo {author} {\bibfnamefont
			{M.}~\bibnamefont {Sawicki}}, \bibinfo {author} {\bibfnamefont {S.~G.}\
			\bibnamefont {Ebbinghaus}}, \ and\ \bibinfo {author} {\bibfnamefont
			{G.}~\bibnamefont {Schmidt}},\ }\href {\doibase 10.1038/srep20827} {\bibfield
		{journal} {\bibinfo  {journal} {Scientific Reports}\ }\textbf {\bibinfo
			{volume} {6}},\ \bibinfo {pages} {20827} (\bibinfo {year}
		{2016})}\BibitemShut {NoStop}%
	\bibitem [{\citenamefont {Evelt}\ \emph {et~al.}(2016)\citenamefont {Evelt},
		\citenamefont {Demidov}, \citenamefont {Bessonov}, \citenamefont
		{Demokritov}, \citenamefont {Prieto}, \citenamefont {Mu{\~{n}}oz},
		\citenamefont {{Ben Youssef}}, \citenamefont {Naletov}, \citenamefont
		{de~Loubens}, \citenamefont {Klein}, \citenamefont {Collet}, \citenamefont
		{Garcia-Hernandez}, \citenamefont {Bortolotti}, \citenamefont {Cros},\ and\
		\citenamefont {Anane}}]{Evelt2016}%
	\BibitemOpen
	\bibfield  {author} {\bibinfo {author} {\bibfnamefont {M.}~\bibnamefont
			{Evelt}}, \bibinfo {author} {\bibfnamefont {V.~E.}\ \bibnamefont {Demidov}},
		\bibinfo {author} {\bibfnamefont {V.}~\bibnamefont {Bessonov}}, \bibinfo
		{author} {\bibfnamefont {S.~O.}\ \bibnamefont {Demokritov}}, \bibinfo
		{author} {\bibfnamefont {J.~L.}\ \bibnamefont {Prieto}}, \bibinfo {author}
		{\bibfnamefont {M.}~\bibnamefont {Mu{\~{n}}oz}}, \bibinfo {author}
		{\bibfnamefont {J.}~\bibnamefont {{Ben Youssef}}}, \bibinfo {author}
		{\bibfnamefont {V.~V.}\ \bibnamefont {Naletov}}, \bibinfo {author}
		{\bibfnamefont {G.}~\bibnamefont {de~Loubens}}, \bibinfo {author}
		{\bibfnamefont {O.}~\bibnamefont {Klein}}, \bibinfo {author} {\bibfnamefont
			{M.}~\bibnamefont {Collet}}, \bibinfo {author} {\bibfnamefont
			{K.}~\bibnamefont {Garcia-Hernandez}}, \bibinfo {author} {\bibfnamefont
			{P.}~\bibnamefont {Bortolotti}}, \bibinfo {author} {\bibfnamefont
			{V.}~\bibnamefont {Cros}}, \ and\ \bibinfo {author} {\bibfnamefont
			{A.}~\bibnamefont {Anane}},\ }\href {\doibase 10.1063/1.4948252} {\bibfield
		{journal} {\bibinfo  {journal} {Applied Physics Letters}\ }\textbf {\bibinfo
			{volume} {108}},\ \bibinfo {pages} {172406} (\bibinfo {year}
		{2016})}\BibitemShut {NoStop}%
	\bibitem [{\citenamefont {Jungfleisch}\ \emph {et~al.}(2015)\citenamefont
		{Jungfleisch}, \citenamefont {Zhang}, \citenamefont {Jiang}, \citenamefont
		{Chang}, \citenamefont {Sklenar}, \citenamefont {Wu}, \citenamefont
		{Pearson}, \citenamefont {Bhattacharya}, \citenamefont {Ketterson},
		\citenamefont {Wu},\ and\ \citenamefont {Hoffmann}}]{Jungfleisch2015}%
	\BibitemOpen
	\bibfield  {author} {\bibinfo {author} {\bibfnamefont {M.~B.}\ \bibnamefont
			{Jungfleisch}}, \bibinfo {author} {\bibfnamefont {W.}~\bibnamefont {Zhang}},
		\bibinfo {author} {\bibfnamefont {W.}~\bibnamefont {Jiang}}, \bibinfo
		{author} {\bibfnamefont {H.}~\bibnamefont {Chang}}, \bibinfo {author}
		{\bibfnamefont {J.}~\bibnamefont {Sklenar}}, \bibinfo {author} {\bibfnamefont
			{S.~M.}\ \bibnamefont {Wu}}, \bibinfo {author} {\bibfnamefont {J.~E.}\
			\bibnamefont {Pearson}}, \bibinfo {author} {\bibfnamefont {A.}~\bibnamefont
			{Bhattacharya}}, \bibinfo {author} {\bibfnamefont {J.~B.}\ \bibnamefont
			{Ketterson}}, \bibinfo {author} {\bibfnamefont {M.}~\bibnamefont {Wu}}, \
		and\ \bibinfo {author} {\bibfnamefont {A.}~\bibnamefont {Hoffmann}},\ }\href
	{\doibase 10.1063/1.4916027} {\bibfield  {journal} {\bibinfo  {journal}
			{Journal of Applied Physics}\ }\textbf {\bibinfo {volume} {117}},\ \bibinfo
		{pages} {17D128} (\bibinfo {year} {2015})}\BibitemShut {NoStop}%
	\bibitem [{\citenamefont {{Houchen Chang}}\ \emph {et~al.}(2014)\citenamefont
		{{Houchen Chang}}, \citenamefont {{Peng Li}}, \citenamefont {{Wei Zhang}},
		\citenamefont {{Tao Liu}}, \citenamefont {Hoffmann}, \citenamefont
		{{Longjiang Deng}},\ and\ \citenamefont {{Mingzhong Wu}}}]{Chang2014}%
	\BibitemOpen
	\bibfield  {author} {\bibinfo {author} {\bibnamefont {{Houchen Chang}}},
		\bibinfo {author} {\bibnamefont {{Peng Li}}}, \bibinfo {author} {\bibnamefont
			{{Wei Zhang}}}, \bibinfo {author} {\bibnamefont {{Tao Liu}}}, \bibinfo
		{author} {\bibfnamefont {A.}~\bibnamefont {Hoffmann}}, \bibinfo {author}
		{\bibnamefont {{Longjiang Deng}}}, \ and\ \bibinfo {author} {\bibnamefont
			{{Mingzhong Wu}}},\ }\href {\doibase 10.1109/LMAG.2014.2350958} {\bibfield
		{journal} {\bibinfo  {journal} {IEEE Magnetics Letters}\ }\textbf {\bibinfo
			{volume} {5}},\ \bibinfo {pages} {6700104} (\bibinfo {year}
		{2014})}\BibitemShut {NoStop}%
	\bibitem [{\citenamefont {Schoen}\ \emph {et~al.}(2015)\citenamefont {Schoen},
		\citenamefont {Shaw}, \citenamefont {Nembach}, \citenamefont {Weiler},\ and\
		\citenamefont {Silva}}]{Schoen2015}%
	\BibitemOpen
	\bibfield  {author} {\bibinfo {author} {\bibfnamefont {M.~A.~W.}\
			\bibnamefont {Schoen}}, \bibinfo {author} {\bibfnamefont {J.~M.}\
			\bibnamefont {Shaw}}, \bibinfo {author} {\bibfnamefont {H.~T.}\ \bibnamefont
			{Nembach}}, \bibinfo {author} {\bibfnamefont {M.}~\bibnamefont {Weiler}}, \
		and\ \bibinfo {author} {\bibfnamefont {T.~J.}\ \bibnamefont {Silva}},\ }\href
	{\doibase 10.1103/PhysRevB.92.184417} {\bibfield  {journal} {\bibinfo
			{journal} {Physical Review B}\ }\textbf {\bibinfo {volume} {92}},\ \bibinfo
		{pages} {184417} (\bibinfo {year} {2015})}\BibitemShut {NoStop}%
	\bibitem [{\citenamefont {Kalarickal}\ \emph {et~al.}(2006)\citenamefont
		{Kalarickal}, \citenamefont {Krivosik}, \citenamefont {Wu}, \citenamefont
		{Patton}, \citenamefont {Schneider}, \citenamefont {Kabos}, \citenamefont
		{Silva},\ and\ \citenamefont {Nibarger}}]{Kalarickal2006}%
	\BibitemOpen
	\bibfield  {author} {\bibinfo {author} {\bibfnamefont {S.~S.}\ \bibnamefont
			{Kalarickal}}, \bibinfo {author} {\bibfnamefont {P.}~\bibnamefont
			{Krivosik}}, \bibinfo {author} {\bibfnamefont {M.}~\bibnamefont {Wu}},
		\bibinfo {author} {\bibfnamefont {C.~E.}\ \bibnamefont {Patton}}, \bibinfo
		{author} {\bibfnamefont {M.~L.}\ \bibnamefont {Schneider}}, \bibinfo {author}
		{\bibfnamefont {P.}~\bibnamefont {Kabos}}, \bibinfo {author} {\bibfnamefont
			{T.~J.}\ \bibnamefont {Silva}}, \ and\ \bibinfo {author} {\bibfnamefont
			{J.~P.}\ \bibnamefont {Nibarger}},\ }\href {\doibase 10.1063/1.2197087}
	{\bibfield  {journal} {\bibinfo  {journal} {Journal of Applied Physics}\
		}\textbf {\bibinfo {volume} {99}},\ \bibinfo {pages} {093909} (\bibinfo
		{year} {2006})}\BibitemShut {NoStop}%
	\bibitem [{\citenamefont {Sebastian}\ \emph {et~al.}(2015)\citenamefont
		{Sebastian}, \citenamefont {Schultheiss}, \citenamefont {Obry}, \citenamefont
		{Hillebrands},\ and\ \citenamefont {Schultheiss}}]{Sebastian2015}%
	\BibitemOpen
	\bibfield  {author} {\bibinfo {author} {\bibfnamefont {T.}~\bibnamefont
			{Sebastian}}, \bibinfo {author} {\bibfnamefont {K.}~\bibnamefont
			{Schultheiss}}, \bibinfo {author} {\bibfnamefont {B.}~\bibnamefont {Obry}},
		\bibinfo {author} {\bibfnamefont {B.}~\bibnamefont {Hillebrands}}, \ and\
		\bibinfo {author} {\bibfnamefont {H.}~\bibnamefont {Schultheiss}},\ }\href
	{\doibase 10.3389/fphy.2015.00035} {\bibfield  {journal} {\bibinfo  {journal}
			{Frontiers in Physics}\ }\textbf {\bibinfo {volume} {3}},\ \bibinfo {pages}
		{1} (\bibinfo {year} {2015})}\BibitemShut {NoStop}%
	\bibitem [{\citenamefont {Sun}\ \emph {et~al.}(2013)\citenamefont {Sun},
		\citenamefont {Chang}, \citenamefont {Kabatek}, \citenamefont {Song},
		\citenamefont {Wang}, \citenamefont {Jantz}, \citenamefont {Schneider},
		\citenamefont {Wu}, \citenamefont {Montoya}, \citenamefont {Kardasz},
		\citenamefont {Heinrich}, \citenamefont {{Te Velthuis}}, \citenamefont
		{Schultheiss},\ and\ \citenamefont {Hoffmann}}]{Sun2013}%
	\BibitemOpen
	\bibfield  {author} {\bibinfo {author} {\bibfnamefont {Y.}~\bibnamefont
			{Sun}}, \bibinfo {author} {\bibfnamefont {H.}~\bibnamefont {Chang}}, \bibinfo
		{author} {\bibfnamefont {M.}~\bibnamefont {Kabatek}}, \bibinfo {author}
		{\bibfnamefont {Y.~Y.}\ \bibnamefont {Song}}, \bibinfo {author}
		{\bibfnamefont {Z.}~\bibnamefont {Wang}}, \bibinfo {author} {\bibfnamefont
			{M.}~\bibnamefont {Jantz}}, \bibinfo {author} {\bibfnamefont
			{W.}~\bibnamefont {Schneider}}, \bibinfo {author} {\bibfnamefont
			{M.}~\bibnamefont {Wu}}, \bibinfo {author} {\bibfnamefont {E.}~\bibnamefont
			{Montoya}}, \bibinfo {author} {\bibfnamefont {B.}~\bibnamefont {Kardasz}},
		\bibinfo {author} {\bibfnamefont {B.}~\bibnamefont {Heinrich}}, \bibinfo
		{author} {\bibfnamefont {S.~G.}\ \bibnamefont {{Te Velthuis}}}, \bibinfo
		{author} {\bibfnamefont {H.}~\bibnamefont {Schultheiss}}, \ and\ \bibinfo
		{author} {\bibfnamefont {A.}~\bibnamefont {Hoffmann}},\ }\href {\doibase
		10.1103/PhysRevLett.111.106601} {\bibfield  {journal} {\bibinfo  {journal}
			{Physical Review Letters}\ }\textbf {\bibinfo {volume} {111}},\ \bibinfo
		{pages} {1} (\bibinfo {year} {2013})}\BibitemShut {NoStop}%
	\bibitem [{\citenamefont {Hillebrands}\ and\ \citenamefont
		{Ounadjela}(2003)}]{Hillebrands2003}%
	\BibitemOpen
	\bibfield  {author} {\bibinfo {author} {\bibfnamefont {B.}~\bibnamefont
			{Hillebrands}}\ and\ \bibinfo {author} {\bibfnamefont {K.}~\bibnamefont
			{Ounadjela}},\ }\href {\doibase 10.1007/3-540-46097-7} {\emph {\bibinfo
			{title} {Spin Dynamics in Confined Magnetic Structures II}}},\ edited by\
	\bibinfo {editor} {\bibfnamefont {B.}~\bibnamefont {Hillebrands}}\ and\
	\bibinfo {editor} {\bibfnamefont {K.}~\bibnamefont {Ounadjela}},\ \bibinfo
	{series} {Topics in Applied Physics}, Vol.~\bibinfo {volume} {87}\ (\bibinfo
	{publisher} {Springer Berlin Heidelberg},\ \bibinfo {address} {Berlin,
		Heidelberg},\ \bibinfo {year} {2003})\BibitemShut {NoStop}%
	\bibitem [{\citenamefont {Nembach}\ \emph {et~al.}(2011)\citenamefont
		{Nembach}, \citenamefont {Silva}, \citenamefont {Shaw}, \citenamefont
		{Schneider}, \citenamefont {Carey}, \citenamefont {Maat},\ and\ \citenamefont
		{Childress}}]{Nembach2011}%
	\BibitemOpen
	\bibfield  {author} {\bibinfo {author} {\bibfnamefont {H.~T.}\ \bibnamefont
			{Nembach}}, \bibinfo {author} {\bibfnamefont {T.~J.}\ \bibnamefont {Silva}},
		\bibinfo {author} {\bibfnamefont {J.~M.}\ \bibnamefont {Shaw}}, \bibinfo
		{author} {\bibfnamefont {M.~L.}\ \bibnamefont {Schneider}}, \bibinfo {author}
		{\bibfnamefont {M.~J.}\ \bibnamefont {Carey}}, \bibinfo {author}
		{\bibfnamefont {S.}~\bibnamefont {Maat}}, \ and\ \bibinfo {author}
		{\bibfnamefont {J.~R.}\ \bibnamefont {Childress}},\ }\href {\doibase
		10.1103/PhysRevB.84.054424} {\bibfield  {journal} {\bibinfo  {journal}
			{Physical Review B}\ }\textbf {\bibinfo {volume} {84}},\ \bibinfo {pages}
		{054424} (\bibinfo {year} {2011})}\BibitemShut {NoStop}%
	\bibitem [{\citenamefont {Berger}\ \emph {et~al.}(2018)\citenamefont {Berger},
		\citenamefont {Edwards}, \citenamefont {Nembach}, \citenamefont {Karenowska},
		\citenamefont {Weiler},\ and\ \citenamefont {Silva}}]{Berger2018}%
	\BibitemOpen
	\bibfield  {author} {\bibinfo {author} {\bibfnamefont {A.~J.}\ \bibnamefont
			{Berger}}, \bibinfo {author} {\bibfnamefont {E.~R.~J.}\ \bibnamefont
			{Edwards}}, \bibinfo {author} {\bibfnamefont {H.~T.}\ \bibnamefont
			{Nembach}}, \bibinfo {author} {\bibfnamefont {A.~D.}\ \bibnamefont
			{Karenowska}}, \bibinfo {author} {\bibfnamefont {M.}~\bibnamefont {Weiler}},
		\ and\ \bibinfo {author} {\bibfnamefont {T.~J.}\ \bibnamefont {Silva}},\
	}\href {\doibase 10.1103/PhysRevB.97.094407} {\bibfield  {journal} {\bibinfo
			{journal} {Physical Review B}\ }\textbf {\bibinfo {volume} {97}},\ \bibinfo
		{pages} {094407} (\bibinfo {year} {2018})}\BibitemShut {NoStop}%
	\bibitem [{\citenamefont {Kittel}(1948)}]{Kittel1948}%
	\BibitemOpen
	\bibfield  {author} {\bibinfo {author} {\bibfnamefont {C.}~\bibnamefont
			{Kittel}},\ }\href {\doibase 10.1103/PhysRev.73.155} {\bibfield  {journal}
		{\bibinfo  {journal} {Physical Review}\ }\textbf {\bibinfo {volume} {73}},\
		\bibinfo {pages} {155} (\bibinfo {year} {1948})}\BibitemShut {NoStop}%
	\bibitem [{\citenamefont {Woltersdorf}\ \emph {et~al.}(2013)\citenamefont
		{Woltersdorf}, \citenamefont {Hoffmann}, \citenamefont {Bauer},\ and\
		\citenamefont {Back}}]{Woltersdorf2013}%
	\BibitemOpen
	\bibfield  {author} {\bibinfo {author} {\bibfnamefont {G.}~\bibnamefont
			{Woltersdorf}}, \bibinfo {author} {\bibfnamefont {F.}~\bibnamefont
			{Hoffmann}}, \bibinfo {author} {\bibfnamefont {H.~G.}\ \bibnamefont {Bauer}},
		\ and\ \bibinfo {author} {\bibfnamefont {C.~H.}\ \bibnamefont {Back}},\
	}\href {\doibase 10.1103/PhysRevB.87.054422} {\bibfield  {journal} {\bibinfo
			{journal} {Physical Review B}\ }\textbf {\bibinfo {volume} {87}},\ \bibinfo
		{pages} {054422} (\bibinfo {year} {2013})}\BibitemShut {NoStop}%
	\bibitem [{\citenamefont {McMichael}, \citenamefont {Twisselmann},\ and\
		\citenamefont {Kunz}(2003)}]{McMichael2003}%
	\BibitemOpen
	\bibfield  {author} {\bibinfo {author} {\bibfnamefont {R.~D.}\ \bibnamefont
			{McMichael}}, \bibinfo {author} {\bibfnamefont {D.~J.}\ \bibnamefont
			{Twisselmann}}, \ and\ \bibinfo {author} {\bibfnamefont {A.}~\bibnamefont
			{Kunz}},\ }\href {\doibase 10.1103/PhysRevLett.90.227601} {\bibfield
		{journal} {\bibinfo  {journal} {Physical Review Letters}\ }\textbf {\bibinfo
			{volume} {90}},\ \bibinfo {pages} {4} (\bibinfo {year} {2003})}\BibitemShut
	{NoStop}%
	\bibitem [{\citenamefont {Tserkovnyak}, \citenamefont {Brataas},\ and\
		\citenamefont {Bauer}(2002)}]{Tserkovnyak2002}%
	\BibitemOpen
	\bibfield  {author} {\bibinfo {author} {\bibfnamefont {Y.}~\bibnamefont
			{Tserkovnyak}}, \bibinfo {author} {\bibfnamefont {A.}~\bibnamefont
			{Brataas}}, \ and\ \bibinfo {author} {\bibfnamefont {G.~E.~W.}\ \bibnamefont
			{Bauer}},\ }\href {\doibase 10.1103/PhysRevLett.88.117601} {\bibfield
		{journal} {\bibinfo  {journal} {Physical Review Letters}\ }\textbf {\bibinfo
			{volume} {88}},\ \bibinfo {pages} {117601} (\bibinfo {year}
		{2002})}\BibitemShut {NoStop}%
	\bibitem [{\citenamefont {Haertinger}\ \emph {et~al.}(2015)\citenamefont
		{Haertinger}, \citenamefont {Back}, \citenamefont {Lotze}, \citenamefont
		{Weiler}, \citenamefont {Gepr{\"{a}}gs}, \citenamefont {Huebl}, \citenamefont
		{Goennenwein},\ and\ \citenamefont {Woltersdorf}}]{Haertinger2015}%
	\BibitemOpen
	\bibfield  {author} {\bibinfo {author} {\bibfnamefont {M.}~\bibnamefont
			{Haertinger}}, \bibinfo {author} {\bibfnamefont {C.~H.}\ \bibnamefont
			{Back}}, \bibinfo {author} {\bibfnamefont {J.}~\bibnamefont {Lotze}},
		\bibinfo {author} {\bibfnamefont {M.}~\bibnamefont {Weiler}}, \bibinfo
		{author} {\bibfnamefont {S.}~\bibnamefont {Gepr{\"{a}}gs}}, \bibinfo {author}
		{\bibfnamefont {H.}~\bibnamefont {Huebl}}, \bibinfo {author} {\bibfnamefont
			{S.~T.~B.}\ \bibnamefont {Goennenwein}}, \ and\ \bibinfo {author}
		{\bibfnamefont {G.}~\bibnamefont {Woltersdorf}},\ }\href {\doibase
		10.1103/PhysRevB.92.054437} {\bibfield  {journal} {\bibinfo  {journal}
			{Physical Review B}\ }\textbf {\bibinfo {volume} {92}},\ \bibinfo {pages}
		{054437} (\bibinfo {year} {2015})}\BibitemShut {NoStop}%
	\bibitem [{\citenamefont {Brataas}\ \emph {et~al.}(2017)\citenamefont
		{Brataas}, \citenamefont {Tserkovnyak}, \citenamefont {Bauer},\ and\
		\citenamefont {Kelly}}]{Brataas2017}%
	\BibitemOpen
	\bibfield  {author} {\bibinfo {author} {\bibfnamefont {A.}~\bibnamefont
			{Brataas}}, \bibinfo {author} {\bibfnamefont {Y.}~\bibnamefont
			{Tserkovnyak}}, \bibinfo {author} {\bibfnamefont {G.~E.~W.}\ \bibnamefont
			{Bauer}}, \ and\ \bibinfo {author} {\bibfnamefont {P.~J.}\ \bibnamefont
			{Kelly}},\ }\href {\doibase 10.1093/oso/9780198787075.003.0008} {\emph
		{\bibinfo {title} {Spin Current}}},\ Vol.~\bibinfo {volume} {1}\ (\bibinfo
	{publisher} {Oxford University Press},\ \bibinfo {year} {2017})\ pp.\
	\bibinfo {pages} {93--142}\BibitemShut {NoStop}%
	\bibitem [{\citenamefont {Raeburn}\ and\ \citenamefont
		{Aldridge}(1978)}]{Raeburn1978}%
	\BibitemOpen
	\bibfield  {author} {\bibinfo {author} {\bibfnamefont {S.~J.}\ \bibnamefont
			{Raeburn}}\ and\ \bibinfo {author} {\bibfnamefont {R.~V.}\ \bibnamefont
			{Aldridge}},\ }\href {\doibase 10.1088/0305-4608/8/9/014} {\bibfield
		{journal} {\bibinfo  {journal} {Journal of Physics F: Metal Physics}\
		}\textbf {\bibinfo {volume} {8}},\ \bibinfo {pages} {1917} (\bibinfo {year}
		{1978})}\BibitemShut {NoStop}%
	\bibitem [{\citenamefont {{De Vries}}(1988)}]{DeVries1988}%
	\BibitemOpen
	\bibfield  {author} {\bibinfo {author} {\bibfnamefont {J.~W.}\ \bibnamefont
			{{De Vries}}},\ }\href {\doibase 10.1016/0040-6090(88)90478-6} {\bibfield
		{journal} {\bibinfo  {journal} {Thin Solid Films}\ }\textbf {\bibinfo
			{volume} {167}},\ \bibinfo {pages} {25} (\bibinfo {year} {1988})}\BibitemShut
	{NoStop}%
	\bibitem [{\citenamefont {Schoen}\ \emph {et~al.}(2016)\citenamefont {Schoen},
		\citenamefont {Thonig}, \citenamefont {Schneider}, \citenamefont {Silva},
		\citenamefont {Nembach}, \citenamefont {Eriksson}, \citenamefont {Karis},\
		and\ \citenamefont {Shaw}}]{Schoen2016}%
	\BibitemOpen
	\bibfield  {author} {\bibinfo {author} {\bibfnamefont {M.~A.~W.}\
			\bibnamefont {Schoen}}, \bibinfo {author} {\bibfnamefont {D.}~\bibnamefont
			{Thonig}}, \bibinfo {author} {\bibfnamefont {M.~L.}\ \bibnamefont
			{Schneider}}, \bibinfo {author} {\bibfnamefont {T.~J.}\ \bibnamefont
			{Silva}}, \bibinfo {author} {\bibfnamefont {H.~T.}\ \bibnamefont {Nembach}},
		\bibinfo {author} {\bibfnamefont {O.}~\bibnamefont {Eriksson}}, \bibinfo
		{author} {\bibfnamefont {O.}~\bibnamefont {Karis}}, \ and\ \bibinfo {author}
		{\bibfnamefont {J.~M.}\ \bibnamefont {Shaw}},\ }\href {\doibase
		10.1038/nphys3770} {\bibfield  {journal} {\bibinfo  {journal} {Nature
				Physics}\ }\textbf {\bibinfo {volume} {12}},\ \bibinfo {pages} {839}
		(\bibinfo {year} {2016})}\BibitemShut {NoStop}%
	\bibitem [{\citenamefont {Edwards}, \citenamefont {Nembach},\ and\
		\citenamefont {Shaw}(2019)}]{Edwards2019}%
	\BibitemOpen
	\bibfield  {author} {\bibinfo {author} {\bibfnamefont {E.~R.}\ \bibnamefont
			{Edwards}}, \bibinfo {author} {\bibfnamefont {H.~T.}\ \bibnamefont
			{Nembach}}, \ and\ \bibinfo {author} {\bibfnamefont {J.~M.}\ \bibnamefont
			{Shaw}},\ }\href {\doibase 10.1103/PhysRevApplied.11.054036} {\bibfield
		{journal} {\bibinfo  {journal} {Physical Review Applied}\ }\textbf {\bibinfo
			{volume} {11}},\ \bibinfo {pages} {054036} (\bibinfo {year}
		{2019})}\BibitemShut {NoStop}%
	\bibitem [{\citenamefont {Shaw}\ \emph {et~al.}(2013)\citenamefont {Shaw},
		\citenamefont {Nembach}, \citenamefont {Silva},\ and\ \citenamefont
		{Boone}}]{Shaw2013a}%
	\BibitemOpen
	\bibfield  {author} {\bibinfo {author} {\bibfnamefont {J.~M.}\ \bibnamefont
			{Shaw}}, \bibinfo {author} {\bibfnamefont {H.~T.}\ \bibnamefont {Nembach}},
		\bibinfo {author} {\bibfnamefont {T.~J.}\ \bibnamefont {Silva}}, \ and\
		\bibinfo {author} {\bibfnamefont {C.~T.}\ \bibnamefont {Boone}},\ }\href
	{\doibase 10.1063/1.4852415} {\bibfield  {journal} {\bibinfo  {journal}
			{Journal of Applied Physics}\ }\textbf {\bibinfo {volume} {114}},\ \bibinfo
		{pages} {243906} (\bibinfo {year} {2013})}\BibitemShut {NoStop}%
	\bibitem [{\citenamefont {Arias}\ and\ \citenamefont
		{Mills}(1999)}]{Arias1999}%
	\BibitemOpen
	\bibfield  {author} {\bibinfo {author} {\bibfnamefont {R.}~\bibnamefont
			{Arias}}\ and\ \bibinfo {author} {\bibfnamefont {D.~L.}\ \bibnamefont
			{Mills}},\ }\href {\doibase 10.1103/PhysRevB.60.7395} {\bibfield  {journal}
		{\bibinfo  {journal} {Physical Review B}\ }\textbf {\bibinfo {volume} {60}},\
		\bibinfo {pages} {7395} (\bibinfo {year} {1999})}\BibitemShut {NoStop}%
	\bibitem [{\citenamefont {Chen}\ \emph {et~al.}(2018)\citenamefont {Chen},
		\citenamefont {Mankovsky}, \citenamefont {Wimmer}, \citenamefont {Schoen},
		\citenamefont {K{\"{o}}rner}, \citenamefont {Kronseder}, \citenamefont
		{Schuh}, \citenamefont {Bougeard}, \citenamefont {Ebert}, \citenamefont
		{Weiss},\ and\ \citenamefont {Back}}]{Chen2018}%
	\BibitemOpen
	\bibfield  {author} {\bibinfo {author} {\bibfnamefont {L.}~\bibnamefont
			{Chen}}, \bibinfo {author} {\bibfnamefont {S.}~\bibnamefont {Mankovsky}},
		\bibinfo {author} {\bibfnamefont {S.}~\bibnamefont {Wimmer}}, \bibinfo
		{author} {\bibfnamefont {M.~A.~W.}\ \bibnamefont {Schoen}}, \bibinfo {author}
		{\bibfnamefont {H.~S.}\ \bibnamefont {K{\"{o}}rner}}, \bibinfo {author}
		{\bibfnamefont {M.}~\bibnamefont {Kronseder}}, \bibinfo {author}
		{\bibfnamefont {D.}~\bibnamefont {Schuh}}, \bibinfo {author} {\bibfnamefont
			{D.}~\bibnamefont {Bougeard}}, \bibinfo {author} {\bibfnamefont
			{H.}~\bibnamefont {Ebert}}, \bibinfo {author} {\bibfnamefont
			{D.}~\bibnamefont {Weiss}}, \ and\ \bibinfo {author} {\bibfnamefont {C.~H.}\
			\bibnamefont {Back}},\ }\href {\doibase 10.1038/s41567-018-0053-8} {\bibfield
		{journal} {\bibinfo  {journal} {Nature Physics}\ }\textbf {\bibinfo {volume}
			{14}},\ \bibinfo {pages} {490} (\bibinfo {year} {2018})}\BibitemShut
	{NoStop}%
	\bibitem [{\citenamefont {Seib}, \citenamefont {Steiauf},\ and\ \citenamefont
		{F{\"{a}}hnle}(2009)}]{Seib2009}%
	\BibitemOpen
	\bibfield  {author} {\bibinfo {author} {\bibfnamefont {J.}~\bibnamefont
			{Seib}}, \bibinfo {author} {\bibfnamefont {D.}~\bibnamefont {Steiauf}}, \
		and\ \bibinfo {author} {\bibfnamefont {M.}~\bibnamefont {F{\"{a}}hnle}},\
	}\href {\doibase 10.1103/PhysRevB.79.092418} {\bibfield  {journal} {\bibinfo
			{journal} {Physical Review B}\ }\textbf {\bibinfo {volume} {79}},\ \bibinfo
		{pages} {092418} (\bibinfo {year} {2009})}\BibitemShut {NoStop}%
	\bibitem [{\citenamefont {Steiauf}\ and\ \citenamefont
		{F{\"{a}}hnle}(2005)}]{Steiauf2005}%
	\BibitemOpen
	\bibfield  {author} {\bibinfo {author} {\bibfnamefont {D.}~\bibnamefont
			{Steiauf}}\ and\ \bibinfo {author} {\bibfnamefont {M.}~\bibnamefont
			{F{\"{a}}hnle}},\ }\href {\doibase 10.1103/PhysRevB.72.064450} {\bibfield
		{journal} {\bibinfo  {journal} {Physical Review B}\ }\textbf {\bibinfo
			{volume} {72}},\ \bibinfo {pages} {064450} (\bibinfo {year}
		{2005})}\BibitemShut {NoStop}%
	\bibitem [{\citenamefont {Safonov}(2002)}]{Safonov2002}%
	\BibitemOpen
	\bibfield  {author} {\bibinfo {author} {\bibfnamefont {V.~L.}\ \bibnamefont
			{Safonov}},\ }\href {\doibase 10.1063/1.1448794} {\bibfield  {journal}
		{\bibinfo  {journal} {Journal of Applied Physics}\ }\textbf {\bibinfo
			{volume} {91}},\ \bibinfo {pages} {8653} (\bibinfo {year}
		{2002})}\BibitemShut {NoStop}%
	\bibitem [{\citenamefont {Pirro}\ \emph {et~al.}(2011)\citenamefont {Pirro},
		\citenamefont {Br{\"{a}}cher}, \citenamefont {Vogt}, \citenamefont {Obry},
		\citenamefont {Schultheiss}, \citenamefont {Leven},\ and\ \citenamefont
		{Hillebrands}}]{Pirro2011}%
	\BibitemOpen
	\bibfield  {author} {\bibinfo {author} {\bibfnamefont {P.}~\bibnamefont
			{Pirro}}, \bibinfo {author} {\bibfnamefont {T.}~\bibnamefont
			{Br{\"{a}}cher}}, \bibinfo {author} {\bibfnamefont {K.}~\bibnamefont {Vogt}},
		\bibinfo {author} {\bibfnamefont {B.}~\bibnamefont {Obry}}, \bibinfo {author}
		{\bibfnamefont {H.}~\bibnamefont {Schultheiss}}, \bibinfo {author}
		{\bibfnamefont {B.}~\bibnamefont {Leven}}, \ and\ \bibinfo {author}
		{\bibfnamefont {B.}~\bibnamefont {Hillebrands}},\ }\href {\doibase
		10.1002/pssb.201147093} {\bibfield  {journal} {\bibinfo  {journal} {Physica
				Status Solidi (B) Basic Research}\ }\textbf {\bibinfo {volume} {248}},\
		\bibinfo {pages} {2404} (\bibinfo {year} {2011})}\BibitemShut {NoStop}%
	\bibitem [{\citenamefont {Demidov}\ \emph {et~al.}(2009)\citenamefont
		{Demidov}, \citenamefont {Kostylev}, \citenamefont {Rott}, \citenamefont
		{Krzysteczko}, \citenamefont {Reiss},\ and\ \citenamefont
		{Demokritov}}]{Demidov2009}%
	\BibitemOpen
	\bibfield  {author} {\bibinfo {author} {\bibfnamefont {V.~E.}\ \bibnamefont
			{Demidov}}, \bibinfo {author} {\bibfnamefont {M.~P.}\ \bibnamefont
			{Kostylev}}, \bibinfo {author} {\bibfnamefont {K.}~\bibnamefont {Rott}},
		\bibinfo {author} {\bibfnamefont {P.}~\bibnamefont {Krzysteczko}}, \bibinfo
		{author} {\bibfnamefont {G.}~\bibnamefont {Reiss}}, \ and\ \bibinfo {author}
		{\bibfnamefont {S.~O.}\ \bibnamefont {Demokritov}},\ }\href {\doibase
		10.1063/1.3231875} {\bibfield  {journal} {\bibinfo  {journal} {Applied
				Physics Letters}\ }\textbf {\bibinfo {volume} {95}},\ \bibinfo {pages}
		{112509} (\bibinfo {year} {2009})}\BibitemShut {NoStop}%
	\bibitem [{\citenamefont {Clausen}\ \emph {et~al.}(2011)\citenamefont
		{Clausen}, \citenamefont {Vogt}, \citenamefont {Schultheiss}, \citenamefont
		{Sch{\"{a}}fer}, \citenamefont {Obry}, \citenamefont {Wolf}, \citenamefont
		{Pirro}, \citenamefont {Leven},\ and\ \citenamefont
		{Hillebrands}}]{Clausen2011}%
	\BibitemOpen
	\bibfield  {author} {\bibinfo {author} {\bibfnamefont {P.}~\bibnamefont
			{Clausen}}, \bibinfo {author} {\bibfnamefont {K.}~\bibnamefont {Vogt}},
		\bibinfo {author} {\bibfnamefont {H.}~\bibnamefont {Schultheiss}}, \bibinfo
		{author} {\bibfnamefont {S.}~\bibnamefont {Sch{\"{a}}fer}}, \bibinfo {author}
		{\bibfnamefont {B.}~\bibnamefont {Obry}}, \bibinfo {author} {\bibfnamefont
			{G.}~\bibnamefont {Wolf}}, \bibinfo {author} {\bibfnamefont {P.}~\bibnamefont
			{Pirro}}, \bibinfo {author} {\bibfnamefont {B.}~\bibnamefont {Leven}}, \ and\
		\bibinfo {author} {\bibfnamefont {B.}~\bibnamefont {Hillebrands}},\ }\href
	{\doibase 10.1063/1.3650256} {\bibfield  {journal} {\bibinfo  {journal}
			{Applied Physics Letters}\ }\textbf {\bibinfo {volume} {99}},\ \bibinfo
		{pages} {162505} (\bibinfo {year} {2011})}\BibitemShut {NoStop}%
	\bibitem [{\citenamefont {Demidov}\ \emph {et~al.}(2016)\citenamefont
		{Demidov}, \citenamefont {Urazhdin}, \citenamefont {Liu}, \citenamefont
		{Divinskiy}, \citenamefont {Telegin},\ and\ \citenamefont
		{Demokritov}}]{Demidov2016}%
	\BibitemOpen
	\bibfield  {author} {\bibinfo {author} {\bibfnamefont {V.~E.}\ \bibnamefont
			{Demidov}}, \bibinfo {author} {\bibfnamefont {S.}~\bibnamefont {Urazhdin}},
		\bibinfo {author} {\bibfnamefont {R.}~\bibnamefont {Liu}}, \bibinfo {author}
		{\bibfnamefont {B.}~\bibnamefont {Divinskiy}}, \bibinfo {author}
		{\bibfnamefont {A.}~\bibnamefont {Telegin}}, \ and\ \bibinfo {author}
		{\bibfnamefont {S.~O.}\ \bibnamefont {Demokritov}},\ }\href {\doibase
		10.1038/ncomms10446} {\bibfield  {journal} {\bibinfo  {journal} {Nature
				Communications}\ }\textbf {\bibinfo {volume} {7}},\ \bibinfo {pages} {10446}
		(\bibinfo {year} {2016})}\BibitemShut {NoStop}%
	\bibitem [{\citenamefont {Yamanoi}\ \emph {et~al.}(2013)\citenamefont
		{Yamanoi}, \citenamefont {Yakata}, \citenamefont {Kimura},\ and\
		\citenamefont {Manago}}]{Yamanoi2013}%
	\BibitemOpen
	\bibfield  {author} {\bibinfo {author} {\bibfnamefont {K.}~\bibnamefont
			{Yamanoi}}, \bibinfo {author} {\bibfnamefont {S.}~\bibnamefont {Yakata}},
		\bibinfo {author} {\bibfnamefont {T.}~\bibnamefont {Kimura}}, \ and\ \bibinfo
		{author} {\bibfnamefont {T.}~\bibnamefont {Manago}},\ }\href {\doibase
		10.7567/JJAP.52.083001} {\bibfield  {journal} {\bibinfo  {journal} {Japanese
				Journal of Applied Physics}\ }\textbf {\bibinfo {volume} {52}},\ \bibinfo
		{pages} {083001} (\bibinfo {year} {2013})}\BibitemShut {NoStop}%
	\bibitem [{\citenamefont {Kalinikos}\ and\ \citenamefont
		{Slavin}(1986)}]{Kalinikos1986}%
	\BibitemOpen
	\bibfield  {author} {\bibinfo {author} {\bibfnamefont {B.~A.}\ \bibnamefont
			{Kalinikos}}\ and\ \bibinfo {author} {\bibfnamefont {A.~N.}\ \bibnamefont
			{Slavin}},\ }\href {\doibase 10.1088/0022-3719/19/35/014} {\bibfield
		{journal} {\bibinfo  {journal} {Journal of Physics C: Solid State Physics}\
		}\textbf {\bibinfo {volume} {19}},\ \bibinfo {pages} {7013} (\bibinfo {year}
		{1986})}\BibitemShut {NoStop}%
	\bibitem [{\citenamefont {Liensberger}\ \emph {et~al.}(2019)\citenamefont
		{Liensberger}, \citenamefont {Flacke}, \citenamefont {Rogerson},
		\citenamefont {Althammer}, \citenamefont {Gross},\ and\ \citenamefont
		{Weiler}}]{Liensberger2019}%
	\BibitemOpen
	\bibfield  {author} {\bibinfo {author} {\bibfnamefont {L.}~\bibnamefont
			{Liensberger}}, \bibinfo {author} {\bibfnamefont {L.}~\bibnamefont {Flacke}},
		\bibinfo {author} {\bibfnamefont {D.}~\bibnamefont {Rogerson}}, \bibinfo
		{author} {\bibfnamefont {M.}~\bibnamefont {Althammer}}, \bibinfo {author}
		{\bibfnamefont {R.}~\bibnamefont {Gross}}, \ and\ \bibinfo {author}
		{\bibfnamefont {M.}~\bibnamefont {Weiler}},\ }\href {\doibase
		10.1109/LMAG.2019.2918755} {\bibfield  {journal} {\bibinfo  {journal} {IEEE
				Magnetics Letters}\ }\textbf {\bibinfo {volume} {10}},\ \bibinfo {pages}
		{5503905} (\bibinfo {year} {2019})}\BibitemShut {NoStop}%
	\bibitem [{\citenamefont {Duan}\ \emph {et~al.}(2015)\citenamefont {Duan},
		\citenamefont {Krivorotov}, \citenamefont {Arias}, \citenamefont {Reckers},
		\citenamefont {Stienen},\ and\ \citenamefont {Lindner}}]{Duan2015}%
	\BibitemOpen
	\bibfield  {author} {\bibinfo {author} {\bibfnamefont {Z.}~\bibnamefont
			{Duan}}, \bibinfo {author} {\bibfnamefont {I.~N.}\ \bibnamefont
			{Krivorotov}}, \bibinfo {author} {\bibfnamefont {R.~E.}\ \bibnamefont
			{Arias}}, \bibinfo {author} {\bibfnamefont {N.}~\bibnamefont {Reckers}},
		\bibinfo {author} {\bibfnamefont {S.}~\bibnamefont {Stienen}}, \ and\
		\bibinfo {author} {\bibfnamefont {J.}~\bibnamefont {Lindner}},\ }\href
	{\doibase 10.1103/PhysRevB.92.104424} {\bibfield  {journal} {\bibinfo
			{journal} {Physical Review B}\ }\textbf {\bibinfo {volume} {92}},\ \bibinfo
		{pages} {104424} (\bibinfo {year} {2015})}\BibitemShut {NoStop}%
	\bibitem [{\citenamefont {Stancil}\ and\ \citenamefont
		{Prabhakar}(2009)}]{Stancil2009}%
	\BibitemOpen
	\bibfield  {author} {\bibinfo {author} {\bibfnamefont {D.~D.}\ \bibnamefont
			{Stancil}}\ and\ \bibinfo {author} {\bibfnamefont {A.}~\bibnamefont
			{Prabhakar}},\ }\href@noop {} {\emph {\bibinfo {title} {{Spin Waves}}}}\
	(\bibinfo  {publisher} {Springer, Boston, Ma},\ \bibinfo {year} {2009})\ p.\
	\bibinfo {pages} {332}\BibitemShut {NoStop}%
	\bibitem [{\citenamefont {Osborn}(1945)}]{Osborn1945}%
	\BibitemOpen
	\bibfield  {author} {\bibinfo {author} {\bibfnamefont {J.~A.}\ \bibnamefont
			{Osborn}},\ }\href {\doibase 10.1103/PhysRev.67.351} {\bibfield  {journal}
		{\bibinfo  {journal} {Physical Review}\ }\textbf {\bibinfo {volume} {67}},\
		\bibinfo {pages} {351} (\bibinfo {year} {1945})}\BibitemShut {NoStop}%
\end{thebibliography}
\end{document}


\title[]{Supplementary Information: High Spin-Wave Propagation Length Consistent with Low Damping in a Metallic Ferromagnet}

\author{Luis Flacke}
\email{luis.flacke@wmi.badw.de}
\affiliation{\mbox{Walther-Meißner Institute, Bayerische Akademie der Wissenschaften, 85748 Garching, Germany}}
\affiliation{Physics Department, Technical University of Munich, 85748 Garching, Germany}

\author{Lukas Liensberger}
\affiliation{\mbox{Walther-Meißner Institute, Bayerische Akademie der Wissenschaften, 85748 Garching, Germany}}
\affiliation{Physics Department, Technical University of Munich, 85748 Garching, Germany}

\author{Matthias Althammer}
\affiliation{\mbox{Walther-Meißner Institute, Bayerische Akademie der Wissenschaften, 85748 Garching, Germany}}
\affiliation{Physics Department, Technical University of Munich, 85748 Garching, Germany}

\author{Hans Huebl}
\affiliation{\mbox{Walther-Meißner Institute, Bayerische Akademie der Wissenschaften, 85748 Garching, Germany}}
\affiliation{Physics Department, Technical University of Munich, 85748 Garching, Germany}
\affiliation{Nanosystems Initiative Munich, 80799 Munich, Germany}
\affiliation{\mbox{Munich Center for Quantum Science and Technology (MCQST), 80799 Munich, Germany}}

\author{Stephan Geprägs}
\affiliation{\mbox{Walther-Meißner Institute, Bayerische Akademie der Wissenschaften, 85748 Garching, Germany}}

\author{Katrin Schultheiss}
\affiliation{Helmholtz-Zentrum Dresden-Rossendorf, 01328 Dresden, Germany}

\author{Aleksandr Buzdakov}
\affiliation{Helmholtz-Zentrum Dresden-Rossendorf, 01328 Dresden, Germany}

\author{Tobias Hula}
\affiliation{Helmholtz-Zentrum Dresden-Rossendorf, 01328 Dresden, Germany}

\author{Helmut Schultheiss}
\affiliation{Helmholtz-Zentrum Dresden-Rossendorf, 01328 Dresden, Germany}

\author{Eric R. J. Edwards}
\affiliation{\mbox{Quantum Electromagnetics Division, National Institute of Standards and Technology, Boulder, CO 80305, USA}}

\author{Hans T. Nembach}
\affiliation{\mbox{Quantum Electromagnetics Division, National Institute of Standards and Technology, Boulder, CO 80305, USA}}

\author{Justin M. Shaw}
\affiliation{\mbox{Quantum Electromagnetics Division, National Institute of Standards and Technology, Boulder, CO 80305, USA}}

\author{Rudolf Gross}
\affiliation{\mbox{Walther-Meißner Institute, Bayerische Akademie der Wissenschaften, 85748 Garching, Germany}}
\affiliation{Physics Department, Technical University of Munich, 85748 Garching, Germany}
\affiliation{Nanosystems Initiative Munich, 80799 Munich, Germany}
\affiliation{\mbox{Munich Center for Quantum Science and Technology (MCQST), 80799 Munich, Germany}}

\author{Mathias Weiler}
\email{mathias.weiler@wmi.badw.de}
\affiliation{\mbox{Walther-Meißner Institute, Bayerische Akademie der Wissenschaften, 85748 Garching, Germany}}
\affiliation{Physics Department, Technical University of Munich, 85748 Garching, Germany}

\date{\today}

\maketitle

\onecolumn

\section{Saturation Magnetization}

\begin{figure}[h]
	\centering
	\includegraphics{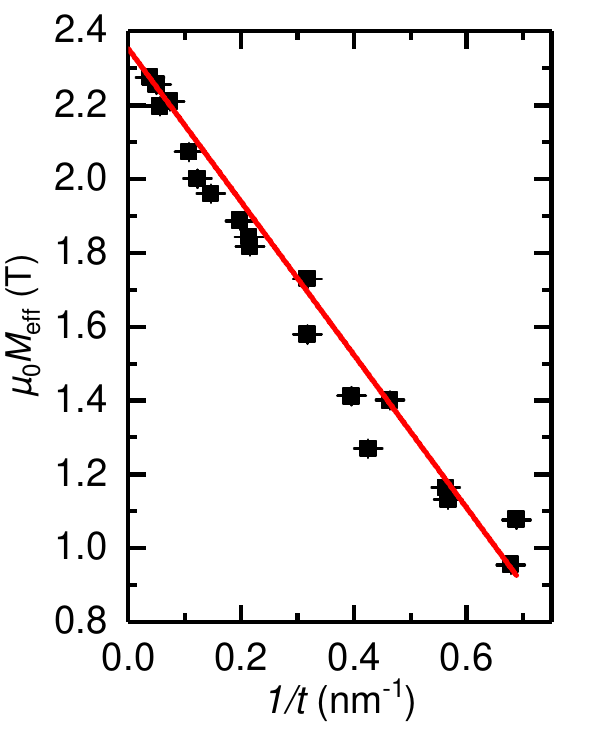}
	\caption{Effective magnetization vs. reciprocal thickness $t$ of the samples. The extrapolated bulk value returns the saturation magnetization $\mu_{\mathrm{0}}M_\mathrm{s}= 2.35$\,T.}
	\label{Meff}
\end{figure}

In order to extract the saturation magnetization of our CoFe samples, we measured the effective magnetization $M_\mathrm{eff}$ and plotted those values vs. the inverse FM thickness (see Fig.\,\ref{Meff}). The y-intercept of a linear fit to the data returns $\mu_{\mathrm{0}}M_\mathrm{s}= (2.35 \pm 0.02)$\,T. This method was shown to result in good agreement with SQUID magnetometry measurements in Schoen's \textit{et al} work~\cite{Schoen2016}.

\newpage
\section{Analysis Procedure for Spin Pumping}

\begin{figure}[h]
	\centering
	\includegraphics{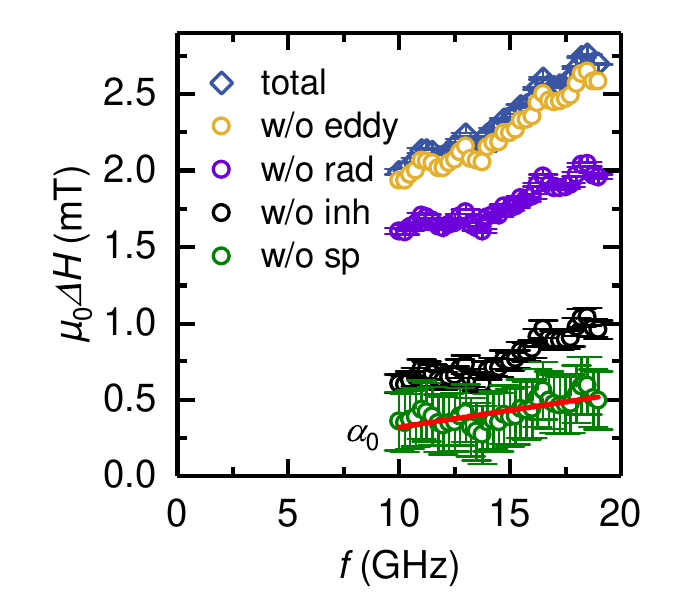}
	\caption{Linewidth vs. frequency for a  Ta(3\,nm)/Al(3\,nm)/\CF(26\,nm)/ Al(3\,nm)/Ta(3\,nm) sample measured by OOP BB-FMR. The blue diamonds show the total measured linewidths, whereas yellow, purple, black and green circles were obtained by substracting different contributions one by one, namely, eddy currents, radiative damping, inhomogeneous linewidth broadening, and spin pumping, respectively.}
	\label{linewidth}
\end{figure}

The total linewidth of the resonance comprises several contributions, coming from different damping processes and long range order inhomogeneities. The total damping $\alpha_{\mathrm{G}} = \alpha_{\mathrm{0}} + \alpha_{\mathrm{sp}} + \alpha_{\mathrm{eddy}} + \alpha_{\mathrm{rad}}$ can be translated to different contributions to the linewidth given as
\begin{equation}\label{dHs}
\mu_{\mathrm{0}}\,\Delta H = \mu_{\mathrm{0}}\,H_{\mathrm{inh}} + 2 \cdot \frac{2\pi f \alpha_\mathrm{G}}{\gamma},
\end{equation}
which is demonstrated for the $t=26\,\mathrm{nm}$ sample in Fig\,\ref{linewidth}. From the total linewidth we extract $\alpha_{\mathrm{G}}$ for the samples as shown in Fig.\,2\,(b). The radiative damping and eddy current contributions to the Gilbert damping parameter are quantified and calculated back to a corresponding linewidth contribution, with help of the second summand of Eq.\,\eqref{dHs}. In order to get an intuition of the influence of each contribution we substract them one by one from the total damping. To exemplify the vanishing influence of eddy currents in our samples, we start by substracted this component ($\alpha_{\mathrm{eddy}} \approx 8.5 \times 10^{-5} $) with remaining linewidth shown as yellow circles. The next two steps are substracting the radiative damping ($\alpha_{\mathrm{rad}} = (4.69 \pm 0.09) \times 10^{-4}$) and the inhomogeneous linewidth broadening ($\mu_\mathrm{0} \Delta H = (1.00 \pm 0.06)\,\mathrm{mT}$). The remaining linewidth comes from the intrinsic damping and spin pumping to the adjacent layers. The latter effect is quantified with Fig\,2\,(b) as $\alpha_{\mathrm{sp}} = (3.45 \pm 0.57) \times 10^{-4}$ and its contribution to the linewidth was then also substracted. The slope of the resulting $\Delta H$ vs. $f$ then allows to determine the intrinsic damping of $\alpha_{\mathrm{0}} = (3.18 \pm 0.48) \times 10^{-4}$ for this sample.

\newpage

\section{In Plane BB-FMR Data for Spin-Wave Propagation}

\begin{figure}[h]
	\centering
	\includegraphics{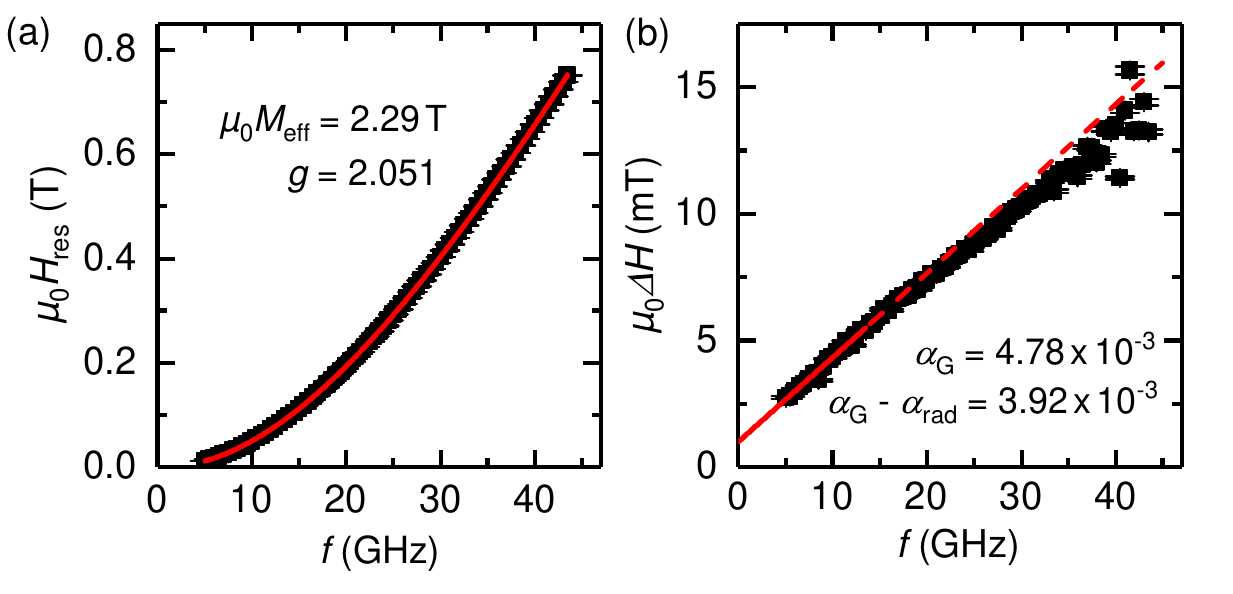}
	\caption{(a) Resonance field vs. frequency. The red curve is a fit to Eq.\,\eqref{IP-Kittel}. (b) Linewidth vs. frequency. The Gilbert damping parameter was extracted from the same frequency range, where we also performed the spin-wave propagation experiments. In order to determine the effective damping for our spin waves in the BLS measurement we substracted the radiative damping contribution which is only present in our BB-FMR setup and not in the patterned devices.}
	\label{RefSample}
\end{figure}

Together with the structured sample B (Pt(3\,nm)/Cu(3\,nm)/\CF(26\,nm)/Cu(3\,nm)/Ta(3\,nm)) we co-deposited blanket films. With these reference samples, we performed in-plane BB-FMR in order to extract the necessary magnetic properties of the structured films to simulate the spin-wave propagation length and dispersion as shown in Fig.\,3\,(f) and (g). From the resonance field $\mu_{\mathrm{0}}H_\mathrm{res}$ vs. $f$ (see Fig.\,\ref{RefSample}), we obtain the Landé-factor $g$ and $\mu_{\mathrm{0}} M_\mathrm{eff}$ by using the in-plane Kittel formula:
\begin{equation}
	\mu_{\mathrm{0}} H_\mathrm{res} = \sqrt{\left(\frac{fh}{g \mu_{\mathrm{B}}}\right)^2 + \left(\frac{\mu_{\mathrm{0}}M_\mathrm{eff}}{2}\right)^2} - \frac{\mu_{\mathrm{0}}M_\mathrm{eff}}{2} -\mu_{\mathrm{0}}H_\mathrm{aniIP}.
	\label{IP-Kittel}
\end{equation}
Here, $h$ is the Planck constant, $\mu_{\mathrm{B}}$ is the Bohr magneton, and $H_\mathrm{aniIP}$ is the growth dependent in-plane anisotropy field, which is negligible in our samples. The difference between $\mu_{\mathrm{0}} M_\mathrm{s}$ and $\mu_{\mathrm{0}} M_\mathrm{eff}$ returns the effective out-of-plane anisotropy field $\mu_{\mathrm{0}} H_\mathrm{k}$. From the slope of the linewidth $\mu_{\mathrm{0}}\Delta H$ vs. $f$, we obtain the damping paramter $\alpha_{\mathrm{G}}$. The radiative damping is substracted as this contribution is not present in our spin-wave propagation experiment. As the BLS measurements was performed in a small frequency range, we only took the low-frequency part ($f<15\,\mathrm{GHz}$) for the determination of $\alpha_{\mathrm{G}} - \alpha_{\mathrm{rad}}$. Note that the slight deviation from linear $\Delta H$ vs. $f$ behavior observed in Fig.\,\ref{RefSample}\,(b) is indicative of two-magnon scattering, as expected in the in-plane configuration.

%